\documentclass[10pt]{article}

\usepackage{amsmath,amssymb,graphicx}
\usepackage{accents}
\usepackage{bbm}
\usepackage{undertilde}
\usepackage{multirow}
\usepackage{tabularx,tabulary,ragged2e,booktabs,caption}
\usepackage[toc,page]{appendix}

\usepackage{color}

\numberwithin{equation}{section}
\numberwithin{figure}{section}
\numberwithin{table}{section}
\allowdisplaybreaks[4]

\definecolor{c20}{rgb}{0.,0,0.}
\definecolor{c30}{rgb}{1,0.5,0.5}
\definecolor{c40}{rgb}{0.3,0.3,0.9}
\definecolor{c50}{rgb}{1,0.9,0.1}

\def\gE#1{\textcolor{c20}{#1}}

\def\aH#1{\textcolor{c20}{#1}}

\newtheorem{theo}{Theorem}[section]
\newtheorem{sat}[theo]{Proposition}
\newtheorem{de}[theo]{Definition}
\newtheorem{lem}[theo]{Lemma}
\newtheorem{exxa}[theo]{Example}
\newtheorem{example}[theo]{Example}
\newtheorem{korr}[theo]{Corollary}

\newtheorem{remarks}[theo]{Remarks}

\newcommand{\proofprop}[1]{ \textbf{Proof of Proposition \ref{#1}}}

\newcommand{\kb}[1]{\boldsymbol{#1}}
\newcommand{\vk}[1]{\kb{#1}}

\newcommand{\E}[1]{\mathbb{E}\left\{#1\right \}}

\newcommand{\pk}[1]{\mbox{\rm$\mathbb{P}$} \left(#1\right) }

\newcommand{\R}{\mathbb{R}}

\newcommand{\N}{\mathbb{N}}

\newcommand{\BQN}{\begin{eqnarray}}
\newcommand{\EQN}{\end{eqnarray}}

\newcommand{\BQNY}{\begin{eqnarray*}}
\newcommand{\EQNY}{\end{eqnarray*}}

\newcommand{\BS}{\begin{sat}}
\newcommand{\ES}{\end{sat}}
\newcommand{\BT}{\begin{theo}}
\newcommand{\ET}{\end{theo}}
\newcommand{\BK}{\begin{korr}}
\newcommand{\EK}{\end{korr}}

\newcommand{\BD}{\begin{de}}
\newcommand{\ED}{\end{de}}

\newcommand{\BIT}{\begin{itemize}}
\newcommand{\EIT}{\end{itemize}}
\newcommand{\BDI}{\begin{description}}
\newcommand{\EDI}{\end{description}}

\newcommand{\BRM}{\begin{remarks}}
\newcommand{\ERM}{\end{remarks}}

\newcommand{\QED}{\hfill $\Box$}

\newcommand{\BTH}{\begin{theo}}
\newcommand{\ETH}{\end{theo}}
\newcommand{\BPR}{\begin{sat}}
\newcommand{\EPR}{\end{sat}}

\newcommand{\BEX}{\begin{exxa}}
\newcommand{\EEX}{\end{exxa}}

\newcommand{\BC}{\begin{cases}}
\newcommand{\EC}{\end{cases}}
\newcommand{\COM}[1]{}
\newcommand{\BL}{\begin{lem}}
\newcommand{\EL}{\end{lem}}

\begin{document}

\begin{center}
	\thispagestyle{empty}

	{\large
	\bf Multivariate Stop loss Mixed Erlang Reinsurance risk: Aggregation, Capital allocation and  Default risk}

    \vskip 0.4 cm

         \centerline{\large      
          Gildas Ratovomirija \footnote{Department of Actuarial Science, 	Faculty of Business and Economics, University of Lausanne, UNIL-Dorigny 1015 Lausanne, Switzerland,} \footnote {
	Vaudoise Assurances, Place de Milan CP 120, 1001 Lausanne,         Switzerland}
}
 \vskip 0.4 cm
 \today{}
\end{center}
{\bf Abstract:} 
In this paper, we address the aggregation of dependent  stop loss reinsurance risks where the dependence among the ceding insurer(s) risks is governed by the Sarmanov distribution and each individual risk belongs to the class of Erlang mixtures. We investigate the effects of the ceding insurer(s) risk dependencies  on the reinsurer risk profile by deriving a closed formula for the distribution function of the aggregated stop loss reinsurance risk. Furthermore,  diversification effects from aggregating reinsurance risks are examined by deriving a closed expression for 
the risk capital  needed for the whole portfolio of the reinsurer and also the allocated risk capital for each business unit under the TVaR capital allocation principle. Moreover, given the risk capital that the reinsurer holds, we express the default probability of the reinsurer analytically. In case the reinsurer is in default, we determine  analytical expressions for the amount of the aggregate reinsured unpaid losses and the unpaid  losses  of each reinsured line of business of the ceding insurer(s). These results are illustrated by numerical examples.   
					\section{Introduction}
Reinsurance companies operate in many regions in the world and insure various insurance business lines. 
In this respect, it is well recognised that the ceding insurer(s) losses are dependent.  This risk dependency can be seen between individual risks within each insurance portfolio and also across  business lines. Furthermore, the phenomena of dependence also occurs from global risk factors which generate  claims simultaneously to each business line,  for instance an hurricane damages buildings or cars which affect property lines, at the same time, causes  people injuries which influence accident lines.
In the risk management framework, for instance the Swiss Solvency Test (SST),  
similarly to insurance companies, reinsurance companies  are obliged to hold a certain level of risk capital in order to be protected from unexpected large losses. 
The determination of this capital requires the aggregation of the losses generated from each reinsurance portfolio whose distribution depends on the  loss distribution of the ceding  insurer(s).
Meyers et al. \cite{meyers2003aggregation} is one of the first contribution
 which have addressed the aggregation of dependent reinsurance risks to evaluate  risk capital. 
In this regard, in order to derive explicit formula for the measure of risk capital including Value-at-Risk (VaR), Tail Value-at-Risk (TVaR)   for the aggregated risk, an important task is the appropriate choice of the marginals and the dependence structure between risks. 
For our framework,  mixed Erlang distribution has been chosen as a claim size model for  the individual risk of the ceding insurer(s).  One of the reason of the tractability of this distribution is the fact that the convolution of such risks belongs again to class of Erlang mixtures, see \cite{Klugman_al08}. Thus stop loss   and  excess of loss premiums  have a closed expression which are very usefull in reinsurance risk modelling, see Lee and Lin \cite {Lee_Lin10,Lee_al12}. In this contribution, we address the dependence structure between risks by the Sarmanov distribution. \\
The aim of this paper is to analyse the effects of the ceding insurer(s) risk dependencies  on the reinsurer risk profile which has only stop loss reinsurance portfolios. Diversification effects from aggregating reinsurance risks are examined by deriving a closed expression for  
the risk capital  needed for the whole portfolio and also the allocated risk capital for each business unit. The effects of the reinsurer default are  also analysed. 
The paper is organised as follows: in Section 2 we describe the background of the Sarmanov distribution as a model for the dependence structure between insurance risk and the mixed Erlang distribution with a common scale parameter as a claim size model. The risk model of the ceding insurer is explored in Section 3, with numerical examples,  by deriving the joint tail probability of the aggregated risk of two portfolios. In Section 4, the aggregation of stop loss mixed Erlang risks of a reinsurer is addressed by determining a closed form for the distribution function (df) of the aggregated risk. Capital allocation and  diversification effects  are also presented with numerical studies.  We also analyse the default risk  of the reinsurer by deriving an analytical form for the expected unpaid losses and the default probability with numerical illustrations.  All the proofs are relegated to Section 5. Some properties of the mixed Erlang distribution are presented in the Appendix. 
					\section{Preliminaries}
			\subsection{Sarmanov distribution}
Due to its flexibility to model the dependence structure between random variables (rv),  the Sarmanov distribution, introduced in  Sarmanov \cite{Sarmanov66}, have been widely used in many fields. Concerning insurance applications, to calulate Bayes premiums in collective risk model Hernandez et al. 
 \cite{hernandez_al12} have addressed   the dependence  between risk profiles using multivariate Sarmanov distribution. Sarabia and Gomez \cite{sarabia_Gomez_11} have used Sarmanov distribution to  fit multivariate insurance count data with Poisson-Beta marginals. The contributions  \cite{Yang_al12, Yang_al13} have explored  tractable asymptotic formulas
 in the context of ruin probabilities where the dependence between insurance risks is governed by the  Saramanov distribution.
\COM{
The Sarmanov distribution introduced in  Sarmanov \cite{Sarmanov66} has proved valuable in numerous insurance applications. For instance
  Hernandez et al. \cite{hernandez_al12} used the multivariate Sarmanov distribution to  model the dependence structure between risk profiles for the calculation of Bayes premiums in the collective risk model. The contribution of Sarabia et Gomez \cite{sarabia_Gomez_11} fitted multivariate insurance count data using the Sarmanov distribution with Poisson-Beta marginals. As shown in  \cite{Yang_al13,Yang_al12}
  the Sarmanov distribution allows for tractable asymptotic formulas
 in the context of ruin probabilities.}
\def\x{\vk{x}}
Refering to \cite{Lee96}, a random vector $(X_1,\ldots,X_n)$ has multivariate Sarmanov distribution with joint density  given by  
\BQN \label{eq:pdfCopulaN}
	h(\x)  = \prod_{i=1}^n  f_i(x_i)  \biggl(
				1+ \sum_{h=2}^{n} \sum_{1 \leqslant j_1 < j_2< \ldots < j_h \leqslant n } 
				\alpha_{ j_1, \ldots ,j_h}
				\prod_{k=1}^h \phi_{j_k}(x_{j_k})
			\biggr), \x:=(x_1,\ldots,x_n),
\EQN
where
$\phi_i$ are  kernel functions,
which are assumed to be bounded and non-constant
 such that
\BQNY \label{eq:CondkernelLee}
\E{\phi_i(X_i)}=0,
\EQNY
\BQN \label{eq:Condition_pdfCopulaN}
				1+ \sum_{h=2}^{n} \sum_{1 \leqslant j_1 < j_2< \ldots < j_h \leqslant n } 
				\alpha_{ j_1, \ldots ,j_h} \prod_{k=1}^h \phi_{j_k}(x_{j_k}) \geqslant 0, \quad \forall x_i \in \R
\EQN
are fulfilled. 
Some general methods for finding the kernel  function $\phi_i$ was specified by Lee \cite{Lee96}  for different types of marginals. In particular,  it is commonly used to choose $\phi_i(x_i)= g_i(x_i)- \E{g_i(X_i)}$ for marginal distributions with support in $\R_{+}$ (see e.g.  \cite{Yang_al13}). The following three cases are the usual specifications of $g_i(x_i)$:\\  
$(i)$  $g_i(x_i)= 2\overline{F}_i(x_i)$ which corresponds to the Farlie-Gumbel-Morgenstern (FGM) distribution, where $\overline{F}_i$ is the survival function of $X_i$,\\ 
	$(ii)$  $g_i(x_i)= x_i^t- \E{X_i^t}$ such that the $t$-th moment $\E{X_i^t} $ of $X_i$ is finite,\\ 
	$(iii)$  $g_i(x_i)= e^{ -t x_i}- \E{e^{ -t X_i}}$ where $\E{e^{ -t X_i}}< \infty$ is the Laplace transform of $X_i$ at $t$.
       \subsection{Mixed Erlang Marginals}
These last decades,  mixed Erlang distribution with a common scale parameter is one of the most usefull model for insurance losses.
In risk theory, using the mixed Erlang distribution as a claim size model, an analytical form for the finite time ruin probability has been derived by Dickson and Willmot \cite{Dickson_al05}  and Dickson \cite{Dickson08}. Recently, using the EM algorithm, mixed Erlang distribution has been fitted  to  catastrophic loss data in the United States by Lee and Lin \cite{Lee_Lin10} and  also to censored and truncated data by Verbelen et al. \cite{verbelenfitting14} .
  Moreover, Lee and Lin \cite{Lee_al12}, Willmot and Woo \cite{Willmot_and_woo_14} have developed the multivariate mixed Erlang distribution to overcome some drawbacks of the copula approach while Badescu et al. \cite{Badescu14} have used multivariate mixed Poisson distribution with mixed Erlang claim sizes to model operational risks. Furthermore, Hashorva and Ratovomirija \cite{Hashorva_Rija14}  have addressed risk aggregation and capital allocation with mixed Erlang marginals and Sarmanov distribution. 
 In the sequel, we denote respectively  
	\BQN\label{pdfErl}
		w_{k}(x,\beta)=
		\frac{\beta^{k} x^{k-1} e^{- \beta x}}{(k-1)!},  \quad
		 W_{k}(x,\beta)=  \sum_{j=k}^{\infty}  \frac{(\beta x )^{j} e^{- \beta x}}{j!}, \quad
		 \overline{W}_{k}(x,\beta)=  \sum_{j=0}^{k-1}  \frac{(\beta x )^{j} e^{- \beta x}}{j!}, x>0,
	\EQN 
the pdf, the df and the survival function of an Erlang distribution	where  $k \in \mathbb{N}^{*}$  is the shape parameter and  $\beta> 0$ is the scale parameter. 
	As its name indicates, the mixed Erlang distribution is elaborated from the Erlang distribution, its pdf and  df are respectively defined as		
\BQN \label{eq:pdfME}
 		f(x,\beta,\utilde{Q})=\sum_ {k=1}^{\infty} q_k w_{k}(x,\beta), \quad
 		F(x,\beta,\utilde{Q})= \sum_ {k=1}^{\infty} q_k W_{k}(x,\beta),
\EQN
	where $\utilde{Q}=(q_1, q_2, \ldots)$ is a vector of non-negative weights satisfying  $\sum_ {k=1}^{\infty} q_k =1$. 
Hereafter we write $X \sim ME(\beta,\utilde{Q})$ if $X$ has pdf and df given by \eqref{eq:pdfME}. 
	  One of the main advantages of the mixed Erlang distribution in insurance risk modeling is the fact that many useful risk related quantities, such as moments and  mean excess function,  have explicit expressions, see e.g., \cite{Lee_Lin10,Willmot_al10, Lee_al12,Willmot_and_woo_14}. 
\COM{
For instance, the quantile function (or VaR) of the mixed Erlang distribution can be easily obtained given the tractable form of the df. From the df in \eqref{eq:pdfME},  at a confidence level $p \in (0,1)$, the VaR of $X$, denoted by $x_p$, is the solution of
	\BQN
		 e^{-\beta x_p} \sum_ {k=1}^{\infty} q_k \sum_ {j=k}^{\infty} \frac{(\beta x_p)^j}{j!} =  p,
	\EQN
which can be solved numerically. 
 Moreover the TVaR of $X$ at a confidence level $p \in (0,1)$ is given by the following explicit formula
	\BQN \label{eq:TVaRME}
		TVaR_X(p)= \frac{1}{1-p} \sum_ {k=1}^{\infty} \frac{k  q_k}{\beta} \overline{W}_{k+1}(x_p,\beta).
	\EQN
}
Furthermore, the mixed Erlang distribution is a tractable marginal distribution for the Sarmanov distribution. Next we present a result for the correlated insurance portfolios.
			\section{ Ceding insurance risk model}
In this section, we  consider two insurance portfolios which both of them consists of $k$ risks and we denote  $S_{1,k} = \sum_{i=1}^k X_i$  and  $S_{2,k} = \sum_{i=k+1}^{2k} X_i$  the aggregated risk of each portfolio   where $X_i,i=1,\ldots, 2k$ is a positive continuous random variable (rv) with finite mean. 
Hereafter, we assume $X_i\sim ME ( \utilde{Q}_i,\beta_i),i=1,\ldots,2k$ and  the dependence structure between risks within and across the portfolio  is governed by the Sarmanov distribution with kernel function 
 $$\phi_i(x_i)= g_i(x_i)- \mathbb{E}(g_i(X_i)),$$
  which shall be abbreviated as 
$$(X_1,\ldots, X_{2k})\sim SME(\vk{\beta}, \utilde{Q})$$  
where $\vk{\beta}=(\beta_1,\ldots, \beta_{2k})$,
 $\utilde{Q}=(\utilde{Q}_1, \ldots, \utilde{Q}_{2k})$.
  In the rest of the paper we consider for \aH{$g_i$ one of}  the three cases described in $(i),(ii)$ and $(iii)$.\\
Furthermore, we define two vectors   of mixing weights $\utilde{\Theta}(\utilde{Q}_i)$ and $\utilde{\Psi}(\utilde{Q}_i)$  where their  components  depend on the kernel function $\phi_i$. In particular,
the components of $\utilde{\Theta}(\utilde{Q}_i)= (\theta_{i,1}, \theta_{i,2}, \ldots)$ are given by:
\BIT
	\item
	for $g_i(x_i)=2\overline{F}(x_i)$,
	$ \theta_{i,s} = \frac{1}{2^{s-1}} 
	\sum_{j=1}^k 
	\begin{pmatrix}
 				s-1\\
				j-1
	\end{pmatrix}
	q_{i,j} \sum_{l=s-j+1}^\infty q_{i,l} , s=1,2,\ldots,$  
	\item 
	for $g_i(x_i)=x_i^t$,
		\BQNY
\theta_{i,s}=
   \left\{
			 	\begin{array}{lcl}
         			0   & \mbox{for} & s \leqslant t , \notag \\
         			\frac{q_{i,s-t}\frac{\Gamma(s)}{\Gamma(s-t)}}{\sum_{j=1}^\infty q_j\frac{\Gamma(j+t)}{\Gamma(j)} }   &\mbox{for} & s > t,
              	\end{array}
        	\right.
\EQNY 
	\item 
	for $ g_i(x_i)=e^{-tx_i}$,
	$ \theta_{i,s} = \frac{q_{i,s} \overline{\beta}^s}{\sum_{j=1}^\infty q_{i,j}\overline{\beta}^j } 
	 $ 
	 with $\overline{\beta} =\frac{\beta}{\beta + t},s=1,2,\ldots,$  
\EIT
whilst  the components of
 $\utilde{\Psi}(\utilde{Q}_i) =(\psi_{i,1}, \psi_{i,2}, \ldots)$ are given by 
\BQNY
\psi_{i,s}
=\sum_ {j=1}^{s} q_{i,j} 	
		 	\begin{pmatrix}
 				k-1\\
				j-1
			\end{pmatrix}		
 		\left(\frac{\beta_i}{Z(\beta_{2k})}\right)^j \left(1-\frac{\beta_i}{Z(\beta_{2k})}
 		\right)^{s-j},
 \EQNY
 where 
 $Z(\beta_{2k})= 2 \beta_{2k}$ for $g_i(x_i)=2\overline{F}(x_i)$,
  $Z(\beta_{2k})=  \beta_{2k}$ for $g_i(x_i)=x_i^t$ and
  $Z(\beta_{2k})=  \beta_{2k}+t$ for $g_i(x_i)=e^{-tx_i}$.
Moreover, for given  mixing weights $\utilde{V}_i=(v_{i1}, v_{i2},\ldots),i=1,\ldots,n+1$ 
we define a vector of mixing probability $\Pi(\utilde{V}_1,\ldots, \utilde{V}_{n+1})$ 
 as follows
\BQNY
		\pi_l \{\utilde{V}_1,\ldots, \utilde{V}_{n+1}\}=
			\left\{
			 	\begin{array}{rcl}
         			0  & \mbox{for} & l=1,\ldots,n, \\
         			\sum_ {j=n}^{l-1} \pi_j\{\utilde{V}_1,\ldots, \utilde{V}_{n}\} v_{n+1,l-j}& 		    \mbox{for} & l=n+1,n+2,\ldots.
              	\end{array}
        	\right.
	\EQNY
We present next the main result of this section. 
 \begin{sat}\label{proposition:survival}
If $(X_1, \ldots, X_{2k})\sim SME(\vk{\beta}, \utilde{Q})$ with $\beta_{2k} \ge \beta_i,i=1,\ldots, 2k-1 $,  then the joint tail probability of $S_{1,k}$ and $S_{2,k}$ is given by (set $\gamma_{j_m} := \E{ g_{j_m}(X_{j_m}) } $)
\BQNY
		\pk{S_{1,k}>u_1,S_{2,k}>u_2}	
			&=&
		 \xi_1\overline{F}_{S_{1,k}}(u_1)
			\overline{F}_{S_{2,k}}(u_2)
		 + \sum_{l=1}^{2k} \sum_{j_1,j_2,\ldots, j_l}  
			\xi_{j_1,j_2,\ldots,j_{l}} \prod_{m=1}^l \gamma_{j_m}
			\gE{\overline{F}_{\tilde{S}_{1,k}} (u_1)
			\overline{F}_{\tilde{S}_{2,k}}(u_2),}
	\EQNY
	where 
      \BQNY
	&&\xi_1 
		=
			 1+ \sum_{j_1} \sum_{j_2 } \alpha_{j_1,j_2} \gamma_{j_1} \gamma_{j_2} 
			- \sum_{j_1} \sum_{j_2 } \sum_{j_3 } \alpha_{j_1,j_2,j_3}  \gamma_{j_1} \gamma_{j_2} \gamma_{j_3}  + \ldots + (-1)^{2k} \alpha_{1,\ldots,2k} \prod_{i=1}^{2k} \gamma_{i}, \notag \\
	&&\xi_{j_1}
		=
			 \sum_{j_1} 
			\biggl(
			- \sum_{j_2 } \alpha_{j_1,j_2}\gamma_{j_2} + 
		 	\sum_{j_2 } \sum_{j_3 } \alpha_{j_1,j_2,j_3}\gamma_{j_2} \gamma_{j_3}   + \ldots +  (-1)^{2k+1} \alpha_{1,\ldots,2k}\prod_{i\in C\backslash \{j_1\}} \gamma_{i}
		 	\biggr), \notag \\
	&& \xi_{j_1,j_2}
		=
			 \sum_{j_1} \sum_{j_2 }
			\biggl(\alpha_{j_1,j_2} -
		  \sum_{j_3 } \alpha_{j_1,j_2,j_3} \gamma_{j_3}+ \sum_{j_3 } \sum_{j_4} \alpha_{j_1,j_2,j_3,j_4} \gamma_{j_3} \gamma_{j_4}  + \ldots + (-1)^{2k} \alpha_{1,\ldots,2k}\prod_{i\in C\backslash \{j_1,j_2\}} \gamma_{i}
		 	\biggr), \notag \\
	&& \xi_{j_1,j_2,j_3}
		=
			\sum_{j_1} \sum_{j_2 }\sum_{j_3 }
			\biggl(\alpha_{j_1,j_2,j_3} - 
		   \sum_{j_4} \alpha_{j_1,j_2,j_3,j_4}  \gamma_{j_4} +   
		   \sum_{j_4 } \sum_{j_5} \alpha_{j_1,j_2,j_3,j_4,j_5}  \gamma_{j_4} \gamma_{j_5}+  \ldots + (-1)^{2k+1} \alpha_{1,\ldots,2k} \prod_{i\in C\backslash \{j_1,j_2,j_3\}} \gamma_{i}
		 	\biggr), \notag \\	 	
	&& \xi_{j_1,\ldots,j_{2k-1}}
	=
		  \sum_{j_1} \ldots \sum_{j_{2k-1} } 
		    \alpha_{j_1,\ldots, j_{2k-1}} - \alpha_{1,\ldots,2k} \gamma_{j_{2k}},
 \notag \\
	&& \xi_{j_1,\ldots,j_{2,k}}=	 \alpha_{1,\ldots,2k}, 		
\EQNY
with
$C=\{1,\ldots,2k \}$,
$j_1 \in C, j_2 \in C\backslash \{j_1\} , j_3 \in C\backslash \{j_1,j_2\}, \ldots, j_{2k} \in C\backslash \{j_1,\ldots,j_{2k-1}\}, $  
\gE{\BQNY   
&&S_{1,k} \sim ME( \Pi \{ \undertilde{\Psi}(\undertilde{Q}_1), \ldots, \undertilde{\Psi}(\undertilde{Q}_k) ,Z(\beta_{2k}) \}),\\
 &&
 S_{2,k} \sim ME( \Pi \{ \undertilde{\Psi}(\undertilde{Q}_{k+1}), \ldots, \undertilde{\Psi}(\undertilde{Q}_{2k}) ,Z(\beta_{2k}) \}),
 \\
	   &&
	   \tilde{S}_{1,k} \sim ME( \Pi \{ \undertilde{\Psi}(\undertilde{Q}_1^{*}), \ldots, \undertilde{\Psi}( \undertilde{Q}_k ^{*}) ,Z(\beta_{2k})\}),
	  \\
	  &&
	  \tilde{S}_{2,k} \sim ME( \Pi \{ \undertilde{\Psi}(\undertilde{Q}_{k+1}^{*}), \ldots, \undertilde{\Psi}( \undertilde{Q}_{2k}^{*}),Z(\beta_{2k}) \}),
\EQNY}
and for  $i=1,\ldots,2k $
$$\gE{\undertilde{Q}_i^*} = 
 		\left\{
			 	\begin{array}{lcl}
         			\undertilde{Q}_i & \mbox{if} & i \notin \{ j_1,j_2,\ldots, j_l \}, \\
         			 \undertilde{\Theta}( \undertilde{Q}_i) & 		    \mbox{if} & i \in \{ j_1,j_2,\ldots, j_l \}.
              	\end{array}
      \right.$$ 	
\end {sat}
\begin{example} \label{ex:marginal param}
Assume that  the ceding insurer has two portfolios say Portfolio A and Portfolio B. Concerning the dependence structure betwen risks, two cases of kernel function are considered    $\phi_i(x_i)=2\overline{F}_i(x_i)-1$ which defines the FGM distribution as explored in  \cite{Cossette_al13} and
$\phi_i(x_i)=e^{-x_i}- \E{e^{-X_i}}$ introduced by Hashorva  and Ratovomirija \cite{Hashorva_Rija14} for mixed Erlang marginals. In the rest of the paper we refer to the latter as the Laplace case.  
Table \ref{table:statX1X2} presents the parameters of each  individual risk $ X_i,i=1,\ldots,4$ and their central moments, whilst Table \ref{table:deparam} displays the dependence parameters between $X_1,X_2,X_3$ and $X_4$. We note that these dependence parameters have been chosen so that   \eqref{eq:Condition_pdfCopulaN} holds.
%
 	%
 	\begin{center}
	\begin{tabular}{|c|c|c| c| c|c|c|c|c|}
		\hline
		& $X_i$ & $\beta_i$ & $\utilde{Q}_i$ & Mean & Variance & Skewness & Kurtosis \\
		
		\hline
		\multirow{2}{*}{Portfolio A} & $X_1$& 0.12 &  (0.4,0.6) & 13.33  & 127.78  &  	 1.55     & 6.50    \\
		
		\cline{2-8} & $X_2$ & 0.14 & (0.3,0.7) & 12.14   &  97.45  &   1.49    &   4.33    \\
			
		\hline
		\hline
		\multirow{2}{*}{Portfolio B} & $X_3$& 0.15 &  (0.5,0.5) & 10.00  & 77.78   &   1.62    &  6.80    \\

	 		\cline{2-8} & $X_4$ & 0.16 & (0.8,0.2)&  7.50  &   53.13  &   1.88    &   8.16   \\
			
		\hline
		\end{tabular}
		\captionof{table}{Parameters and central moments of $X_i,i=1,2,3,4$.} \label{table:statX1X2}
		\end{center}	
\begin{center}
	\begin{tabular}{|c|c|c|c|c|c|c|c|c| c|c| c| }
	
	\hline
	&$\alpha_{1,2}$  &  $\alpha_{1,3}$  & $\alpha_{1,4}$  
	& $\alpha_{2,3}$   & $\alpha_{2,4}$ &	$\alpha_{3,4}$ 
	& $\alpha_{1,2,3}$  &  $\alpha_{1,2,4}$  &    $\alpha_{1,3,4}$  &  
	  $\alpha_{2,3,4}$  &   $\alpha_{1,2,3,4}$   \\
	  
	  \hline
	FGM & 0.6 &  0.1  & 0.1 
	& 0.1  & 0.04 &	0.5 
	& 0.11 &  0.12 &    0.10  &  
	  0.15 &  0.07 \\
	  
	    \hline
	Laplace 
	& 16 &  5  & 3 
	& 5  & 3 & 	8
	& 56 &  30 &    15  &  
	  20 &  170 \\
	  
	\hline
\end{tabular}
		\captionof{table}{ Dependence parameters of $(X_1,X_2,X_3,X_4)$.} \label{table:deparam}
\end{center}	
It can be seen from Table \ref{table:eXcedenceProb} that the  interdependence between the two insurance portfolios yields    high  probability for the aggregated risk of each portfolio to exceed simultaneously some threshold.
\begin{center}
	\begin{tabular}{|c|c|c|c|}
	\hline
	 Thresholds       &   Independence case   & Laplace case & FGM case\\
	\hline
	
	$(u_1,u_2)$ & $ \pk{S_{1,2}>u_1,S_{2,2}>u_2}$ &  $\pk{S_{1,2}>u_1,S_{2,2}>u_2}$ &  $\pk{S_{1,2}>u_1,S_{2,2}>u_2}$  \\
	\hline
	(20,15) &  0.1494 & 0.1569 &0.1573  \\
	
	\hline
	
	(25,20) &  0.0697 &0.0751 &0.0795  \\
	
	\hline
	
	(30,25) &  0.0304 & 0.0331 &0.0374  \\
	
	\hline
	
	(35,30) &  0.0125 & 0.0138 &0.0165 \\
	
	\hline
\end{tabular}
		\captionof{table}{Joint tail probability of $S_{1,2}=X_1+X_2$ and $S_{2,2}=X_3+X_4$.} \label{table:eXcedenceProb}
\end{center}	
\end{example}
			      		\section{Reinsurance risk model}
In this section, we denote $R_2:= T_{1,k} + T_{2,k} $ the aggregate reinsurance stop loss risk   where $T_{1,k}:=(S_{1,k}- d_1 )_+$ and $T_{2,k}:=(S_{2,k}- d_2)_+$ represent two stop loss reinsurance portfolios   with $S_{1,k} = \sum_{i=1}^k X_i$  and  $S_{2,k} = \sum_{i=k+1}^{2k} X_i$  the ceding insurer aggregated risk and $d_i ,i=1,2$ some positive deductible. Additionally,
for a given risk  $ X \sim ME(\beta, \undertilde{Q})$ with df $F$ and for a deductible $d>0$ we denote in the rest of the paper 
$$F_X(y+d)
	= 
	\sum_{k=0}^\infty 
	\Delta_{k}(d,\beta,\utilde{Q} )
	W_{k+1}(y,\beta),$$
$$\overline{U} _X(c,d,\beta)
	= 
	 \sum_ {k=0}^{\infty} 
			 (k+1)   \Delta_k(d,\beta,\utilde{Q})
			 \overline{W}_{k+2}(c,\beta),$$
	with
\BQNY
\Delta_{k} (d,\beta,\utilde{Q}) = \frac{1}{\beta} \sum_{j=0}^\infty 
	q_{j+k+1}w_{j+1}(d,\beta).
\EQNY
Furthermore, for $X_i \sim ME(\utilde{Q}_i,\beta)$, with $d_i>0,i=1,2$ we define  
\BQNY
		&&F_{X_1+X_2}(d_1,d_2,s)= \sum_ {k=0}^{\infty}   \sum_ {j=0}^{\infty}	
					\Delta_k(d_1,\beta,\utilde{Q}_1) \Delta_j(d_2,\beta,\utilde{Q}_2)   W_{k+j+2}(s,\beta),\\
		&&\overline{U}_{X_1}(c,d_1,d_2,\beta)= 
		\frac{1}{\beta}
		\sum_ {k=0}^{\infty}   \sum_ {j=0}^{\infty} (k+1)	
					\Delta_k(d_1,\beta,\utilde{Q}_1) \Delta_j(d_2,\beta,\utilde{Q}_2)  \overline{W}_{k+j+3}(c,\beta),\\
		&&\overline{U}_{X_2}(c,d_1,d_2,\beta)=
		\frac{1}{\beta}
		 \sum_ {k=0}^{\infty}   \sum_ {j=0}^{\infty} (j+1)	
					\Delta_k(d_1,\beta,\utilde{Q}_1) \Delta_j(d_2,\beta,\utilde{Q}_2)   \overline{W}_{k+j+3}(c,\beta),\\
		&&\overline{U}_{X_1+X_2}(c,d_1,d_2,\beta)=
		\frac{1}{\beta}
		 \sum_ {k=0}^{\infty}   \sum_ {j=0}^{\infty} (k+j+2)	
					\Delta_k(d_1,\beta,\utilde{Q}_1) \Delta_j(d_2,\beta,\utilde{Q}_2)   \overline{W}_{k+j+3}(c,\beta).
\EQNY

	\subsection{Aggregation of reinsurance stop loss risks}
In the following result we show that the df of the aggregated stop loss risk $R_2$   has a closed form which allows us to derive analytical formula \aH{of its} mean excess function. 
 \begin{sat}\label{proposition:Aggregation}
If $(X_1, \ldots, X_{2k})\sim SME(\vk{\beta}, \utilde{Q})$ with $\beta_{2k} \ge \beta_i,i=1,\ldots, 2k-1 $  
 and $d_j>0, j=1,2,$   then the df of the aggregated stop loss risk $R_2$ is given by 
 \BQN \label{eq:DF_Stop_Loss}
		F_{R_2}(s)= 
		\left\{
			 	\begin{array}{lcl}
		F_{S_{1,k}, S_{2,k}} (d_1,d_2)
			& \mbox{for} & s=0, \\
	     F_{S_{1,k}, S_{2,k}} (d_1+s,d_2 +s)
			 &\mbox{for} & s>0,
              	\end{array}
        	\right. 	    
	\EQN
where
\BQNY
		F_{S_{1,k}, S_{2,k}} (d_1,d_2) = \xi_1 F_{S_{1,k}}(d_1)
			F_{S_{2,k}}(d_2)
		 + \sum_{l=1}^{2k} \sum_{j_1,j_2,\ldots, j_l}  
			\xi_{j_1,j_2,\ldots,j_{l}} \prod_{m=1}^l \gamma_{j_m}
			F_{\tilde{S}_{1,k}}(d_1)
			F_{\tilde{S}_{2,k}}(d_2),
\EQNY
\gE{
\BQNY
		 F_{S_{1,k}, S_{2,k}} (d_1+s,d_2 +s)
		 &=& 
		 	\xi_1 \biggl(
		 	F_{S_{1,k}}(d_1)F_{S_{2,k}}(d_2+s)+	
		 	F_{S_{1,k}}(d_1+s)F_{S_{2,k}}(d_2)	+
		 	F_{S_{1,k}+S_{2,k}}(d_1,d_2,s)
		 	\biggr) \\
		 	&& 
		 	+ \sum_{l=1}^{2k} \sum_{j_1,j_2,\ldots, j_l}  
			\xi_{j_1,j_2,\ldots,j_{l}} \prod_{m=1}^l \gamma_{j_m} 
			\biggl(
			F_{\tilde{S}_{1,k}}(d_1)
			F_{\tilde{S}_{2,k}}(d_2+s)
			\\
		 	&& +
		 	F_{\tilde{S}_{1,k}}(d_1+s)
			F_{\tilde{S}_{2,k}}(d_2)
			+ F_{\tilde{S}_{1,k}+\tilde{S}_{2,k}}(d_1,d_2,s)
			\biggr) ,
\EQNY}
with $S_{1,k},S_{2,k},\tilde{S}_{1,k},\tilde{S}_{2,k}$
are
 defined  in Proposition \ref{proposition:survival}.
\BRM \label{rem:risk measures Aggregate}
Given the tractable form of the df in \eqref{eq:DF_Stop_Loss}, many risk related quantities for $R_2$ have an explicit form, for instance, for $c>0$ the mean excess function of $R_2$ is given by 
\gE{\BQN  \label{eq: Meanexcess_R2}
		\E{R_2-c \vert R_2>c}
			&=& \frac{1}{ \overline{F}_{R_2}(c)} 
			\biggl[
			\xi_1 \biggl(
		 	F_{S_{1,k}}(d_1) \overline{U}_{S_{2,k}}(c,d_2,Z(\beta_{2k}))+	
		 	F_{S_{2,k}}(d_2)	\overline{U}_{S_{1,k}}(c,d_1,Z(\beta_{2k}))
		 	\notag\\ 
		 	&& +
		 	\overline{U}_{S_{1,k}+S_{2,k} }(c,d_1,d_2,Z(\beta_{2k}))
		 	\biggr) 
		 	+ \sum_{l=1}^{2k} \sum_{j_1,j_2,\ldots, j_l}  
			\xi_{j_1,j_2,\ldots,j_{l}} \prod_{m=1}^l \gamma_{j_m} 
			\biggl(
			F_{\tilde{S}_{1,k}}(d_1)
			\overline{U}_{\tilde{S}_{2,k}}(c,d_2,Z(\beta_{2k}))
			\notag\\ 
		 	&& +
		 	F_{\tilde{S}_{2,k}}(d_2)
		 	\overline{U}_{\tilde{S}_{1,k}}(c,d_1,Z(\beta_{2k}))
			+ \overline{U}_{\tilde{S}_{1,k}+\tilde{S}_{2,k}}(c,d_1,d_2,Z(\beta_{2k}))
			\biggr) 
			 \biggr] - c.
\EQN}
\ERM
\end{sat}
\begin{example} \label{ex:reinsurer}
In this illustration, we consider the same parameters of each individual risk of the ceding insurer portfolios as in Table \ref{table:statX1X2}. Furthermore, we assume that the ceding insurer re-insures its two portfolios to a reinsurer with  stop loss programs where the deductibles are  $d_1=40$ and $d_2=30$ for Portfolio A and for Portfolio B, respectively. 
In practice, it is recognised that risk measures on the aggregated risk are sensitive to the strength of the dependence between individual risks. Actually, by taking into account the dependence within and accross the ceding insurer portfolios which is determined by the parameters in Table \ref{table:deparam}, the aggregated risk $R_2$ of the reinsurer is  riskier than in the independence case. Therefore, based on VaR and TVaR as a risk measure, the reinsurer needs much more risk capital in the dependence case. Furthermore, for a different confidence level $p$, it can be seen that the deviation from the independence assumption is greater for VaR than for TVaR.   
\begin{center}
	\begin{tabular}{||c| | c|c||c|c||c|c||}
	\hline
Confidence level		 &\multicolumn{2}{c||}{Independence case}  & \multicolumn{2}{c||}{Laplace case} & \multicolumn{2}{c||}{FGM case} \\
	\hline
	   p  & $VaR_{R_2}(p)$ &  $TVaR_{R_2}(p)$  & $VaR_{R_2}(p)$ &  $TVaR_{R_2}(p)$   & $VaR_{R_2}(p)$ &  $TVaR_{R_2}(p)$  \\
	\hline
		90.00 \%  &  11.73  &   22.64 & 11.98     &  22.93    & 13.92 & 25.35    \\
		
		\hline
		92.50 \% &  14.98 &   25.76  & 15.24	&   26.06   &   17.36  & 28.62  \\
		
		\hline
		95.00 \% & 19.47  &  30.10  &  19.75	&  30.41   &  22.10  & 33.14  \\
		
		\hline
		97.50 \% & 26.97 &   37.40  &   27.27 	&  37.73    &  29.93  &  40.68  \\
		
		\hline
		99.00 \%  & 36.64  &  46.85  &  36.97   	&  47.21   &  39.93  & 50.40 \\
		
		\hline
		99.50 \% & 43.80  &  53.89 &  44.15   	&   54.25    &  47.30  & 57.59  \\

		\hline
		99.90 \% & 60.08  & 69.92  &  60.45    	&   70.31    &  63.91  & 73.89 \\

		\hline
 \end{tabular}
		\captionof{table}{Deviation of VaR and TVaR from the independence case.} \label{table:risk measure}
\end{center}	
It is well known that risk diversification across portfolios arises from aggregating their individual risks, see e.g., \cite{tasche2005measuring}, \cite{tang2006economic}. In this respect, by considering the TVaR as a measure for the risk capital,  the diversification benefits $D_p(T_1,T_2)$  are quantified as the relative reduction of the risk capital required for the whole portfolio of the reinsurer from aggregating  the stop loss risk $T_1$ and $T_2$ as follows  
				$$ D_p(T_1,T_2) = 1- \frac{TVaR_{R_2}(p)}{TVaR_{T_1}(p) + TVaR_{T_2}(p)}.$$ 		%
As presented in Table \ref{table:diversification}, diversification benefits increase with the confidence level. 
Conversely, the deviation from the independence case yields a reduction of the diversification benefits which is obvious since the full diversification effects are attained when risks are independent.   
\begin{center}
	\begin{tabular}{|c|c| c|c|c|c|}
		\hline

		& p &  $TVaR_{R_2}(p)$&    $TVaR_{T_1}(p)$  & $TVaR_{T_2}(p) $  & $D_p(T_1,T_2)$ (\%)  \\

		\hline
		
		\multirow{4}{*}{Independence case} & 95.00 \% &  30.10 	& 24.87 	&  18.26  & 30.19  \\

		\cline{2- 6} & 97.50 \% &  37.40 & 32.34 	&  24.26 	&  33.92  \\
		
		\cline{2- 6} & 	99.00 \%   &  46.85  & 41.89 	&  31.97   & 36.56   \\

		\cline{2- 6}&	99.90 \% &  69.92   &  64.84		&  50.55   &  39.40   \\
		
		\hline
		\hline

		\multirow{4}{*}{Laplace case} & 95.00 \% &  30.41     &   25.11     & 18.41	   &  30.13   \\

		\cline{2- 6} & 97.50 \%    &   37.73   &   32.58  	&  24.42    &  33.82   \\
		
		\cline{2- 6} &	99.00 \%   &   47.21    &   42.14   	&    32.13   &  36.44   \\

		\cline{2- 6} &  99.90 \%   &    70.31    &   65.07   	 &   50.72    &   39.28 \\
		
		\hline
		\hline
		\multirow{4}{*}{FGM case} & 95.00 \% & 33.14 &  27.71 &	 18.64   & 28.52   \\

		\cline{2- 6} & 97.50 \%    &  40.68  &   35.40 	&  24.69    &  32.29   \\
		
		\cline{2- 6} &	99.00 \%   & 50.40   &   45.14 	&  32.44    &  35.03   \\

		\cline{2- 6} &  99.90 \%   &    73.89    &      68.32   	 &      51.08     & 38.11   \\
		
		\hline
	
		\end{tabular}
		\captionof{table}{ Diversification benefits based on TVaR of the aggregate risk $R_2$ and the individual risk $T_i, i=1,2$.} \label{table:diversification}
\end{center}
\end{example}
		\subsection{TVaR capital allocation}
In this section, we derive analytical expressions for the amount of capital allocated to each individual risk of the reinsurer under the TVaR  principle.
In the enterprise risk management framework, to absorb large unexpected losses,  reinsurer are required to hold a certain amount of  economic capital for the entire portfolio.  In this respect, the so-called capital allocation consists in attributing the required capital to each individual line. This allows the reinsurance company to identify and to monitor efficiently their risks. In the literature, many capital allocation techniques have been  developed, see for instance\cite{cummins2000allocation, tasche2004allocating,tasche2005measuring, Dhaene_et_al12,Mcneil_al05} and references therein. In practice, it is well known that the TVaR  principle takes into account the dependence structure between risks and satisfy the full allocation principle. More precisely, if $R_n=\sum_{i=1}^n T_i$ is the aggregate risk where $T_i$ is a  rv with finite mean that represents the individual risk of the reinsurer, the amount of capital $TVaR_{p}(T_i,R_n)$ required for each risk $T_i$, for $i=1,\ldots,n,$ is defined as 

\BQN\label{eq:TVaR_Alloc_Def}
   TVaR_{p}(T_i,R_n) = \frac{\mathbb{E}(T_i \mathbbm{1}_{\{R_n > VaR_{R_n}(p)\}})}{1-p},
\EQN 
where $p \in (0,1) $ is the tolerence level.
The full allocation principle implies
 $$TVaR_{R_n}(p)=\sum_{i=1}^n  TVaR_{p}(T_i,R_n)$$
 which means that, based on TVaR as a risk measure,  the capital required for the entire portfolio is equal to the sum of the required capital of each risk within the portfolio.
The following proposition develops an explicit form for  $TVaR_{p}(T_i,R_2), i=1,2$, in the case of   stop loss mixed Erlang type risks.
In addition,   we define below $S_{1,k},S_{2,k},\tilde{S}_{1,k},\tilde{S}_{2,k}$  as in Proposition \ref{proposition:Aggregation} and we denote  $x_p:=  VaR_{R_2}(p)$.
 \begin{sat}\label{proposition:TVaRCapital}
Let $(X_1,\ldots, X_{2k})\sim SME(\vk{\beta}, \utilde{Q})$ with $\beta_{2k} \ge \beta_i,i=1,\ldots, 2k-1 $  
 and $d_j>0, j=1,2$. If further $T_j,  j=1,2$ has finite mean   then 
\gE{\BQNY
  TVaR_{p}(T_1,R_2)&=&
	\frac{1}{ 1-p} 
			\biggl[
			\xi_1 \biggl(
		 	F_{S_{2,k}}(d_2)	\overline{U}_{S_{1,k}}(x_p,d_1,Z(\beta_{2k}))+
		 	\overline{U}_{S_{1,k}}(x_p,d_1,d_2,Z(\beta_{2k}))
		 	\biggr) \notag\\ 
		 	&& 
		 	+ \sum_{l=1}^{2k} \sum_{j_1,j_2,\ldots, j_l}  
			\xi_{j_1,j_2,\ldots,j_{l}} \prod_{m=1}^l \gamma_{j_m} 
			\biggl(
		 	F_{\tilde{S}_{2,k}}(d_2)
		 	\overline{U}_{\tilde{S}_{1,k}}(x_p,d_1,Z(\beta_{2k}))
		 	\notag\\ 
		 	&& 
			+ \overline{U}_{\tilde{S}_{1,k}}(x_p,d_1,d_2,Z(\beta_{2k}))
			\biggr) 
			 \biggr].
\EQNY   }
\end{sat}
\begin{example}
In this example, we consider the same individual risks  and dependence parameters as in Example \ref{ex:marginal param} and the reinsurance programs as in Example  \ref{ex:reinsurer}. Based on $TVaR$  as a risk measure for quantifying the risk capital required for the whole portfolio 
 the required capital of each stop loss risk $T_i,i=1,2$ 
  are evaluated for different confidence level $p$. Since $T_1$ is riskier than $T_2$, as displayed in  Table \ref{table:Exactrisk_measures_Dependences}, more capital is required for $T_1$ compared to the amount needed for  $T_2$. 
\begin{center}
	\begin{tabular}{||c||c| c|c||c|c| c||}
	
		\hline

		  &   \multicolumn{3}{c||}{Laplace case}& \multicolumn{3}{c||}{FGM case} \\  
		  
		\hline

		 p &  $TVaR_{R_2}(p)$&    $TVaR_{p}(T_1,R_2)$  & $TVaR_{p}(T_2,R_2) $
		  & $TVaR_{R_2}(p)$& $TVaR_{p}(T_1,R_2)$  & $TVaR_{p}(T_2,R_2) $   \\

		\hline
		
		90.00 \% &  22.93  	& 14.56  &   8.37 & 25.35	 &  16.19   &  9.16  \\
		
		\hline
		
		92.50 \% &  26.06  &  16.85   &   9.21  & 28.62  & 18.62   &  10.00  \\
		
		\hline
		
		 95.00 \% & 30.41   &  20.15     &  10.26   & 33.14	  & 22.12  &  11.02   \\

		\hline

		97.50 \%  &  37.73    &  25.99   &  11.74 	& 40.68 	 	 &   28.21  &  12.47   \\
		
		\hline
	
	 	99.00 \%   &  47.21 	&  33.94   & 13.27  & 50.40	    	& 36.39    &  14.01 \\

		\hline
		
		99.50 \%   &  54.25 	&   40.03    &  14.22 	& 57.59 	&  42.59	  &  15.00   \\

		\hline
		
		99.90 \%   &  70.31  	&    54.20    &   16.11 	&  73.89 	&  56.79	  &  17.10   \\

		\hline

		\end{tabular}
		\captionof{table}{  TVaR and allocated capital to each stop loss risk $T_i, i=1,2,$ under the TVaR  capital allocation principle.} \label{table:Exactrisk_measures_Dependences}
\end{center}
\end{example}
		\subsection{ Reinsurer default  analysis}
In the enterprise risk management framework, reinsurers are obliged to hold a certain amount of capital $K>0$ in order to be covered from  unexpected large losses. The amount of this capital is determined so that the reinsurer will be able to honor its liabilities even in the worst case with high probability. For instance, in the SST, $K$ is quantified as the TVaR at a tolerance level $99\%$ of the aggregated  risk $R_n= \sum_{i=1}^n T_i $ where $T_i$ represents the individual risk of the reinsurer.  This means that for $99\%$ probability the reinsurer has enough buffer to pay its obligations. However, in case $R_n>K$ the reinsurer is in default and thus ceding insurers are not protected from losses exceeding $K$ i.e. $R_n-K$. By analogy to the case between the insurer and the policyholders, see \cite{myers2001capital},   the quantity $(R_n-K)_+$ is called the \textit{ default option of the reinsurer} or in other words  the \textit{ceding insurers deficit} with $U(K):=\E{(R_n-K)_+}$ the value of the default option. In view of the full capital allocation principle,  for a given risk capital  $K$ required for the entire portfolio of the reinsurer, if $K_i, i=1, \ldots, n $ is the risk capital needed for each individual risk  then $K=\sum_{i=1}^n K_i$. Furthermore, the value of the default option is also defined as  the sum of the value of the unpaid losses $U(K_i,K):=\E{(T_i- K_i) \mathbbm{1}_{ \{ R_n> K\} } }$ of each ceding insurer(s)  reinsured lines of business, specifically (see e.g. \cite{Dhaene_et_al12}) 
 $$U(K) = \sum_{i=1}^n  U(K_i,K)$$
Next, we determine   $U(K)$ and also derive  an explicit formula for the  default probability of the reinsurer $\phi(K):=\mathbb{P}(R_2>K)$ where $R_2= T_1+T_2$. Furthermore, given that the reinsurer is in default, analytical expressions of the unpaid excess losses of each line of business  of the ceding insurer(s) are derived.
 \begin{sat}\label{proposition:Value of the default option}
If $(X_1, \ldots, X_{2k})\sim SME(\vk{\beta}, \utilde{Q})$ with $\beta_{2k} \ge \beta_i,i=1,\ldots, 2k-1 $  and $d_j>0, j=1,2$  then for a given risk capital $K>0$ 
\gE{
\BQNY
	U(K)&=& 
			\xi_1 \biggl(
		 	F_{S_{1,k}}(d_1) \overline{U}_{S_{2,k}}(K,d_2,Z(\beta_{2k}))+	
		 	F_{S_{2,k}}(d_2)	\overline{U}_{S_{1,k}}(K,d_1,Z(\beta_{2k}))
		 	\notag\\ 
		 	&& +
		 	\overline{U}_{S_{1,k}+S_{2,k} }(K,d_1,d_2,Z(\beta_{2k}))
		 	\biggr) 
		 	+ \sum_{l=1}^{2k} \sum_{j_1,j_2,\ldots, j_l}  
			\xi_{j_1,j_2,\ldots,j_{l}} \prod_{m=1}^l \gamma_{j_m} 
			\biggl(
			F_{\tilde{S}_{1,k}}(d_1)
			\overline{U}_{\tilde{S}_{2,k}}(K,d_2,Z(\beta_{2k}))
			\notag\\ 
		 	&& +
		 	F_{\tilde{S}_{2,k}}(d_2)
		 	\overline{U}_{\tilde{S}_{1,k}}(K,d_1,Z(\beta_{2k}))
			+ \overline{U}_{\tilde{S}_{1,k}+\tilde{S}_{2,k}}(K,d_1,d_2,Z(\beta_{2k}))
			\biggr) 
		 - K \overline{F}_{R_2}(K),
\EQNY}
where $ {\overline{F}_{R_2} (K)} = 1- F_{R_2} (K) $ with $F_{R_2} (.)$ is defined in Proposition \ref{proposition:Aggregation}.
\BRM
In view of Proposition \ref{proposition:Aggregation} analytical expression for the default probability of the reinsurer is given  by 
\gE{
\BQN \label{eq:defaultprob}
	\phi(K) &=& 1-   \xi_1 \biggl(
		 	F_{S_{1,k}}(d_1)F_{S_{2,k}}(d_2+K)+	
		 	F_{S_{1,k}}(d_1+s)F_{S_{2,k}}(d_2)	+
		 	F_{S_{1,k}+S_{2,k}}(d_1,d_2,K)
		 	\biggr) \notag \\
		 	&& 
		 	+ \sum_{l=1}^{2k} \sum_{j_1,j_2,\ldots, j_l}  
			\xi_{j_1,j_2,\ldots,j_{l}} \prod_{m=1}^l \gamma_{j_m} 
			\biggl(
			F_{\tilde{S}_{1,k}}(d_1)
			F_{\tilde{S}_{2,k}}(d_2+K)
			\notag \\
		 	&& +
		 	F_{\tilde{S}_{1,k}}(d_1+K)
			F_{\tilde{S}_{2,k}}(d_2)
			+ F_{\tilde{S}_{1,k}+\tilde{S}_{2,k}}(d_1,d_2,K)
			\biggr).
\EQN}
\ERM		
\end{sat}
 \begin{sat}\label{proposition:unpaid excess losses}
		Let $K_i, i=1,2$ be the capital required for each stop loss reinsurance portfolio of the reinsurer such that $K=K_1+K_2$. Given that the reinsurer is in default, if $(X_1, \ldots, X_{2k})\sim SME(\vk{\beta}, \utilde{Q})$ with $\beta_{2k} \ge \beta_i,i=1,\ldots, 2k-1 $,  $d_j>0, j=1,2$  and  $T_j,  j=1,2$ has finite mean, then 
 \gE{
 \BQNY
		U(K_1,K)&=&
   \xi_1 \biggl(
		 	F_{S_{2,k}}(d_2)	\overline{U}_{S_{1,k}}(K,d_1,Z(\beta_{2k}))+
		 	\overline{U}_{S_{1,k}}(K,d_1,d_2,Z(\beta_{2k}))
		 	\biggr) \notag\\ 
		 	&& 
		 	+ \sum_{l=1}^{2k} \sum_{j_1,j_2,\ldots, j_l}  
			\xi_{j_1,j_2,\ldots,j_{l}} \prod_{m=1}^l \gamma_{j_m} 
			\biggl(
		 	F_{\tilde{S}_{2,k}}(d_2)
		 	\overline{U}_{\tilde{S}_{1,k}}(K,d_1,Z(\beta_{2k}))
		 	\notag\\ 
		 	&& 
			+ \overline{U}_{\tilde{S}_{1,k}}(K,d_1,d_2,Z(\beta_{2k}))
			\biggr) -
			K_1 \overline{F}_{R_2}(K).
	\EQNY }
\end{sat}
\begin{example}
		Consider the required capital for the entire portfolio $K$ as  $TVaR_{R2}(p)$ and for the individual risk $K_i$ as $TVaR_{p}(T_i,R_2),i=1,2$ presented in Table \ref{table:Exactrisk_measures_Dependences}. From Table \ref{table:capital} we can see that an increase of the confidence level $p$ yields an uprise of the capital required for the entire portfolio of the reinsurer which in turn decreases the default probability and also the value of the unpaid losses of each  portfolio of the ceding insurer. 
\begin{center}
	\begin{tabular}{|c|c|| c|c|c||c|c||c|c|}
	
	\hline
	    & p  &  K  &  $\phi(K)$ &  $U(K)$ &  $K_1$   & $U(K_1,K)$ &   $K_2$   & $U(K_2,K)$   \\
	\hline
		\multirow{4}{*}{Independence  case} 
		& 95.00 \%  &   30.10     & 0.01860
   & 0.19288
 &    19.69 	&  0.15436 
   &  10.41  &  0.03852 
  \\

		\cline{2- 9} &  97.50 \% &  37.40    & 0.00929
   & 0.09483
  &  25.47  	&   0.07928 
     &   11.93    &  0.01555 
  \\

		\cline{2- 9} &  99.00 \%  &  46.85   & 0.00370
   &  0.03725
 &  33.35 	 &    0.03228 
  &   13.50 & 0.00497 
  \\

		\cline{2- 9} &  99.90 \% &  69.92    & 0.00036
  & 0.00360
  &  53.59  	&   0.00321 
    &   16.33 &   0.00039 
   \\

		\hline
		\hline
		
		\multirow{4}{*}{Laplace case} 
		& 95.00 \%  &   30.41    & 0.01863
   & 0.19338
 &  20.15  	&   0.15586  
   &  10.26  &    0.03752  
 \\

		\cline{2- 9} &  97.50 \% &   37.73   & 0.00930
   & 0.09750
  & 25.99   	&   0.07979  
      &   11.73    &   0.01525  
  \\

		\cline{2- 9} &  99.00 \%  &  47.21   &  0.00371
  & 0.03731 
  & 	 33.94  &     0.03237   
   &   13.27 &    0.00494 

 \\

		\cline{2- 9} &  99.90 \% &  70.31    & 0.00037
  & 0.00360
  &  54.20  	&   0.00320 
    &   16.11 &    0.00040
  \\

		\hline
		 \hline
	\multirow{4}{*}{FGM case} 
		& 95.00 \%  &   33.14   &  0.01870  & 0.19924  &  22.11	&   0.16237    &  11.02   &  0.03687\\

		\cline{2- 9} &  97.50 \%  &   40.68     &  0.00932  & 0.09740  &    28.21 	&      0.08212     &   12.47    &   0.01528  \\

		\cline{2- 9} &  99.00 \%  & 50.40  &  0.00372  &  0.03805 & 36.39 	 &  0.03286   & 14.01 & 0.00519 \\

		\cline{2- 9} &  99.90 \% &    73.89   & 0.00037  & 0.00364  &  	 56.80 &    0.00317    &  17.10  &  0.00047     \\

		\hline
		
 \end{tabular}
		\captionof{table}{Default probability, default value option of the reinsurer and unpaid losses of the insurer.} \label{table:capital}
\end{center}	
\end{example}
\COM{
			\section{Applications to insurance data}
In this section, applications of the main results in real insurance data are given. In this regards, loss data from an insurance company has been used where the insurer has  two business lines namely the vehicle liability line and  the  CASCO line which consists of the collision portfolio and the natural damage portfolio. Hereafter, we denote $X_1, X_2, X_3$ respectively the aggregate loss from the liability line, the collision and the natural damage portfolio. The aggregated loss  generated by the CASCO lines is defined by $S_2=X_2+X_3$. 
			\subsection{EM algorithm}
 For further details  on EM algorithm, we refer to Lee and Lin (2010, 2012), Badescu et al.(2014) and Verbelen et al.(2014).....  

			\subsection{Dependence parameters estimation}

}
				     \section{Proofs}
\proofprop{proposition:survival}	
%
The joint tail probability of $(S_{1,k} , S_{2,k})$ is determined in terms of the joint density of $(X_1,\ldots,X_{2k})$ as follows
\BQN \label{eq:defN1}
			\mathbb{P}(S_{1,k}>u_1,S_{2,k}>u_2)=
	 		\int \ldots \int_{s_{1,k}>u_1,s_{2,k}>u_2}  h(\x)dx_1\ldots dx_{2k}.
\EQN
\\
Refering to \eqref{eq:pdfCopulaN}, the joint density of  $(X_1,\ldots,X_{2k})$ is given by (set $\gamma_{j_h}:=\E{g_{j_h}(X_{j_h})}$)
\BQNY
	h(\x)&=& 
			\prod_{i=1}^{2k} f_i(x_i) 
			\biggl(
				1+ \sum_{h=2}^{2k} \sum_{1 \leqslant j_1 < j_2<\ldots < j_h \leqslant 2k } 
				\prod_{i=1}^h (g_{j_h}(x_{j_h})- \gamma_{j_h} )
			\biggr)\\
		 &=& 
			\xi_1 \prod_{i=1}^{2k} f_i(x_i) 
			+\xi_{j_1} \tilde{f}_{j_1} (x_{j_1} ) \prod_{i \in C \backslash \{j_1\} } f_i(x_i) 
			+\xi_{j_1,j_2} \tilde{f}_{j_1} (x_{j_1} ) \tilde{f}_{j_2} (x_{j_2} ) 
			\prod_{i \in C \backslash \{j_1,j_2\} } f_i(x_i)\notag \\
			&&+\ldots+
			 \xi_{j_1,\ldots,j_{2k-1}} \prod_{i=1}^{2k-1}\tilde{f}_{j_i} (x_{j_i} ) f_{j_{2k}} (x_{j_{2k}} ) 
			+\xi_{j_1,\ldots,j_{2k}}   \prod_{i=1}^{2k}\tilde{f}_{i} (x_{i} ),
\EQNY
where
$\tilde{f}_{i} (x_{i}) = g(x_i) f_{i} (x_i)$,
\BQNY
	&&\xi_1 
		=
			 1+ \sum_{j_1} \sum_{j_2 } \alpha_{j_1,j_2} \gamma_{j_1} \gamma_{j_2} 
			- \sum_{j_1} \sum_{j_2 } \sum_{j_3 } \alpha_{j_1,j_2,j_3}  \gamma_{j_1} \gamma_{j_2} \gamma_{j_3}  + \ldots + (-1)^{2k} \alpha_{1,\ldots, 2k} \prod_{i=1}^{2k} \gamma_{i}, \notag \\
	&&\xi_{j_1}
		=
			 \sum_{j_1} 
			\biggl(
			- \sum_{j_2 } \alpha_{j_1,j_2}\gamma_{j_2} + 
		 	\sum_{j_2 } \sum_{j_3 } \alpha_{j_1,j_2,j_3}\gamma_{j_2} \gamma_{j_3}   + \ldots +  (-1)^{2k+1} \alpha_{1,\ldots,2k}\prod_{i\in C\backslash \{j_1\}} \gamma_{i}
		 	\biggr), \notag \\
	&& \xi_{j_1,j_2}
		=
			 \sum_{j_1} \sum_{j_2 }
			\biggl(\alpha_{j_1,j_2} -
		  \sum_{j_3 } \alpha_{j_1,j_2,j_3} \gamma_{j_3}+ \sum_{j_3 } \sum_{j_4} \alpha_{j_1,j_2,j_3,j_4} \gamma_{j_3} \gamma_{j_4}  + \ldots + (-1)^{2k} \alpha_{1,\ldots,2k}\prod_{i\in C\backslash \{j_1,j_2\}} \gamma_{i}
		 	\biggr), \notag \\
	&& \xi_{j_1,j_2,j_3}
		=
			\sum_{j_1} \sum_{j_2 }\sum_{j_3 }
			\biggl(\alpha_{j_1,j_2,j_3} - 
		   \sum_{j_4} \alpha_{j_1,j_2,j_3,j_4}  \gamma_{j_4} +   
		   \sum_{j_4 } \sum_{j_5} \alpha_{j_1,j_2,j_3,j_4,j_5}  \gamma_{j_4} \gamma_{j_5}+  \ldots + (-1)^{2k+1} \alpha_{1,\ldots,2k} \prod_{i\in C\backslash \{j_1,j_2,j_3\}} \gamma_{i}
		 	\biggr), \notag \\	 	
	&& \xi_{j_1,\ldots,j_{2k-1}}
	=
		  \sum_{j_1} \ldots \sum_{j_{2k-1} } 
		    \alpha_{j_1,\ldots, j_{2k-1}} - \alpha_{1,\ldots,2k} \gamma_{j_{2k}},
 \notag \\
	&& \xi_{j_1,\ldots,j_{2k}}=	 \alpha_{1,\ldots,2k}, 		
\EQNY
with
$C=\{1,\ldots,2k \}$,
$j_1 \in C, j_2 \in C\backslash \{j_1\} , j_3 \in C\backslash \{j_1,j_2\}, \ldots, j_{2k} \in C\backslash \{j_1,\ldots,j_{2k-1}\}. $\\
After some rearrangements, one can express $h(\x)$  as follows 
\BQNY
	h(\x)
	&=& 
	\xi_1 \prod_{i=1}^{2k} f_i(x_i) + 
	\sum_{l=1}^{2k} \sum_{j_1,j_2,\ldots, j_l}  
	\xi_{j_1,j_2,\ldots,j_{l}}	
	\prod_{h=1}^l \tilde{f}_{j_h}(x_{j_h}) 
	\prod_{i \notin \{j_1,j_2,\ldots, j_l\} }f_i(x_i) \notag \\
	&=& 
	 \xi_1 \prod_{i=1}^{2k} f_i(x_i) + \sum_{l=2}^{2k} \sum_{j_1,j_2,\ldots, j_l}  
	\xi_{j_1,j_2,\ldots,j_{l}} 	
	\prod_{i=1}^{2k} \gE{f_i^{*}(x_i)},
\EQNY
where for $i=1,\ldots,{2k}$
$$\gE{f_i^{*}(x_i)} =  \left\{
			 	   \begin{array}{lcl}
         			 f_i(x_i) & \mbox{if} & i \notin \{ j_1,j_2,\ldots, j_l \}, \\
         			 \tilde{f}_i(x_i) & \mbox{if} & i \in \{j_1,j_2,\ldots, j_l\}.
              	\end{array}
      \right.$$
  By Lemma \ref{lem:tranformation ME}, $\tilde{f}_i$ is a pdf of a mixed Erlang distribution, therefore
     one can write \eqref{eq:defN1} as a sum  product of convolutions of mixed Erlang risks as follows
   \BQN \label{eq:conv}
			&&\mathbb{P}(S_{1,k}>u_1,S_{2,k}>u_2) \notag \\
			&&=
			 \xi_1 \int_{u_1}^{\infty} \int_{u_1-x_{1}}^{\infty}  \ldots\int_ {u_1-x_{1}- \ldots -x_{k-2}}^{\infty} 
	 		 \prod_{i=1} ^{k-1}  f_i(x_i) \overline{F}_{k}(u_1-x_{1}-\ldots - x_{k-1})
			dx_{k-1} \ldots dx_{1}\notag \\
			&&
			\times \int_ {u_2} ^{\infty}  \int_ {u_2-x_{k+1}}^{\infty}  \ldots \int_ {u_2-x_{k+1}- \ldots -x_{2k-2}}^{\infty} 
	 		 \prod_{i=k+1} ^{2k-1}  f_i(x_i) \overline{F}_{k}(u_2-x_{k+1}-\ldots - x_{2k-1})
			dx_{2k-1} \ldots dx_{k+1}\notag \\
			&&+ 
			\sum_{l=1}^{2k} \sum_{j_1,j_2,\ldots, j_l}  
	\xi_{j_1,j_2,\ldots,j_{l}}
	\prod_{m=1}^l \gamma_{j_m} \notag \\
			&&
			\times \int_{u_1}^{\infty} \int_{u_1-x_{1}}^{\infty}  \ldots\int_ {u_1-x_{1}- \ldots -x_{k-2}}^{\infty}
	 		 \prod_{i=1} ^{k-1}  f_i^{*}(x_i) \overline{F}_{k}^{*}(u_1-x_{1}-\ldots - x_{k-1})
			dx_{k-1} \ldots dx_{1}\notag \\
			&&
			\times \int_{u_2} ^{\infty}  \int_ {u_2-x_{k+1}}^{\infty}  \ldots \int_ {u_2-x_{k+1}- \ldots -x_{2k-2}}^{\infty} 
	 		 \prod_{i=k+1} ^{2k-1}  \gE{f_i^{*}(x_i)  \overline{F}_{k}^{*}}(u_2-x_{k+1}-\ldots - x_{2k-1})
			dx_{2k-1} \ldots dx_{k+1}.	
\EQN
\gE{Provided that $\beta_{2k} \ge \beta_i,i=1,\ldots, 2k-1 $, by Lemma \ref{lem:pdfMErlang}  each $i-$th mixed Erlang  component of \eqref{eq:conv}  can be transformed into a new mixed Erlang distribution  with a common scale parameter $Z(\beta_{2k})$.
Therefore, with the help of Remark \ref{rem:AggregationN}, by convolution \eqref{eq:conv} can be expressed as a sum product of two mixed Erlang survival function as follows}
\BQNY
			\mathbb{P}(S_{1,k}>u_1,S_{2,k}>u_2)
			&=&
		 \xi_1\overline{F}_{S_{1,k}}(u_1)
			\overline{F}_{S_{2,k}}(u_2)
		 + \sum_{l=1}^{2k} \sum_{j_1,j_2,\ldots, j_l}  
			\xi_{j_1,j_2,\ldots,j_{l}} \prod_{m=1}^l \gamma_{j_m}
			\gE{\overline{F}_{\tilde{S}_{1,k}}(u_1)
			\overline{F}_{\tilde{S}_{2,k}}(u_2).}
	\EQNY
Thus the proof is complete. 
\QED
\\
\proofprop{proposition:Aggregation}
  Similarly to the independence case described in Lemma \ref{lem:Aggreg_Stop_Loss_Indep}, the df of $R_2$ is of mixed distribution and can be expressed in terms of the joint df of $(T_1,T_2)$ as follows 
	\BQNY 
			F_{R_2}(s)&=&
	       \left\{
			 	\begin{array}{lcl}
         			\pk{T_1 =0 ,  T_2 =0 }   & \mbox{for} & s=0 \notag \\
         			\pk{  T_1 = 0,  0 <T_2  \leqslant s} 
 				   		+ \pk{ 0 < T_1  \leqslant s, T_2 = 0} 
 				   		+ \pk{ T_1 + T_2 \leqslant s , 0< T_1  \leqslant s, 0< T_2  \leqslant s}  &\mbox{for} & s>0
              	\end{array}
        	\right.  
        	\\ 
			&=:&
	       \left\{
			 	\begin{array}{lcl}
         			F_{S_{1,k}, S_{2,k}} (d_1,d_2)   & \mbox{for} & s=0 \\
         			F_{S_{1,k}, S_{2,k}} (d_1,d_2+s)	+ F_{S_{1,k}, S_{2,k}} (d_1+s,d_2) + 
         			F_{S_{1,k}, S_{2,k}}(s+d_1,s+d_2) 
         			&\mbox{for} & s>0.
              	\end{array}
        	\right.                     
	\EQNY
By Proposition \ref{proposition:survival} and Lemma \ref{lem:Aggreg_Stop_Loss_Indep}, $F_{R_2}(s)$ can be written in two terms  as follows: 
\BIT
\item the discrete term  
\BQNY
		F_{S_{1,k}, S_{2,k}} (d_1,d_2) = \xi_1 F_{S_{1,k}}(d_1)
			F_{S_{2,k}}(d_2)
		 + \sum_{l=1}^{2k} \sum_{j_1,j_2,\ldots, j_l}  
			\xi_{j_1,j_2,\ldots,j_{l}} \prod_{m=1}^l \gamma_{j_m}
			F_{\tilde{S}_{1,k}}(d_1)
			F_{\tilde{S}_{2,k}}(d_2),
\EQNY
\item the continuous term 
	\gE{\BQNY
		 &&F_{S_{1,k}, S_{2,k}} (d_1+s,d_2 +s)\\
		 && =
		 	\xi_1 
		 	F_{S_{1,k}}(d_1)F_{S_{2,k}}(d_2+s)
		 	+ \sum_{l=1}^{2k} \sum_{j_1,j_2,\ldots, j_l}  
			\xi_{j_1,j_2,\ldots,j_{l}} \prod_{m=1}^l \gamma_{j_m} 
			F_{\tilde{S}_{1,k}}(d_1)
			F_{\tilde{S}_{2,k}}(d_2+s)
			\\
		 	&& +
		 	\xi_1 F_{S_{1,k}}(d_1+s)F_{S_{2,k}}(d_2)
		 	+
		 	\sum_{l=1}^{2k} \sum_{j_1,j_2,\ldots, j_l}  
			\xi_{j_1,j_2,\ldots,j_{l}} \prod_{m=1}^l \gamma_{j_m} 
		 	F_{\tilde{S}_{1,k}}(d_1+s)
			F_{\tilde{S}_{2k}}(d_2)
			\\
		 	&& 
		 	 +
		 	\xi_1 F_{S_{1,k}+S_{2,k}}(d_1,d_2,s)
		 	+ \sum_{l=1}^{2k} \sum_{j_1,j_2,\ldots, j_l}  
			\xi_{j_1,j_2,\ldots,j_{l}} \prod_{m=1}^l \gamma_{j_m} 
		 	F_{\tilde{S}_{1,k}+\tilde{S}_{2,k}}(d_1,d_2,s).		
\EQNY}
\EIT
 This completes the proof.  
\QED
\\
\proofprop{proposition:TVaRCapital}	
In view of \eqref{eq:TVaR_Alloc_Def}
\BQN \label{eq:defTVaRAlloc1}
   TVaR_{p}(T_1,R_2) = \frac{\mathbb{E}(T_1 \mathbbm{1}_{\{R_2 > VaR_{R_2}(p)\}})}{1-p} 
   =\frac{1}{1-p} \int_{VaR_{R_2}(p)} ^ \infty \mathbb{E} ( T_1\mathbbm{1}_{\{R_2=s\}}) ds.
\EQN
First, we need to calculate 
 $\mathbb{E} ( T_1\mathbbm{1}_{\{R_2 =s\}})$  as follows
  $$\mathbb{E} ( T_1\mathbbm{1}_{\{R_2 =s\}})= \int_{0}^\infty u f_{T_1,T_1+T_2=s}(u) du.$$
Let 
\gE{\BQNY
		&&f_{S_{1,k}+S_{2,k}}(d_1,d_2,u):= \frac{d}{du}F_{S_{1,k}+S_{2,k}}(d_1,d_2,u),\\
		&&f_{\tilde{S}_{1,k}+\tilde{S}_{2,k}}(d_1,d_2,u):=\frac{d}{du}F_{\tilde{S}_{1,k}+\tilde{S}_{2,k}} (d_1,d_2,u).
\EQNY}
As in Proposition \ref{proposition:Aggregation}, one can express $\mathbb{E} ( T_1\mathbbm{1}_{\{R_2 =s\}})$ as follows
\gE{\BQN \label{eq:CondExpT1T}
	\mathbb{E} ( T_1\mathbbm{1}_{\{R_2=s\}}) 
	&=& 
	\xi_1 \biggl(	
		 	F_{S_{2,k}}(d_2)
		 	\int_0 ^{s}  u f_{S_{1,k}}(d_1+u) du	+ 
		 	\int_0 ^{s} u
		 	f_{S_{1,k}+S_{2,k}}(d_1,d_2,u) du 
		 	\biggr) \notag \\
		 	&& 
		 	+ \sum_{l=1}^{2k} \sum_{j_1,j_2,\ldots, j_l}  
			\xi_{j_1,j_2,\ldots,j_{l}} \prod_{m=1}^l \gamma_{j_m} 
			\biggl(
			F_{\tilde{S}_{2,k}}(d_2)
		 	\int_0 ^{s}  u f_{\tilde{S}_{1,k}}(d_1+u) du 
		 	\notag \\
		 	&& 
			+ \int_0 ^{s} u f_{\tilde{S}_{1,k}+\tilde{S}_{2,k} }(d_1,d_2,u) du
			\biggr).
\EQN}
By Lemma \ref{lem:pdfStop_Loss},  for $X_i \sim ME(\beta, \utilde{Q}_i)$  and $d_i >0,i=1,2$
\BQN \label{eq:UX1}
	 \int_0 ^{s}  u f_{X}(d_i+u) du
	=
	\frac{1}{\beta}
	 \sum_ {k=0}^{\infty} 
			 (k+1)   \Delta_k(d_i,\beta,\utilde{Q}_i)
			 \overline{W}_{k+2}(s,\beta)
	=:	\overline{U} _{X_i}(s,d_i,\beta).
\EQN
Similarly, by Lemma \ref{lem:Aggreg_Stop_Loss_Indep} 
\BQN \label{eq:UX1X2}
	 \int_0 ^{s}  u f_{X_1+X_2}(d_1,d_2,u) du
	=
	\frac{1}{\beta}
		 \sum_ {k=0}^{\infty}   \sum_ {j=0}^{\infty} (k+1)	
		\Delta_k(d_1,\beta,\utilde{Q}_1) \Delta_j(d_2,\beta,\utilde{Q}_2)   W_{k+j+3}(s,\beta)
	=:	\overline{U} _{X_1}(s,d_1,d_2,\beta).
\EQN
Taking \eqref{eq:UX1} and \eqref{eq:UX1X2} into account, one may write \eqref{eq:CondExpT1T} as follows
\gE{\BQN \label{eq:CondExpT1TFinal}
	\mathbb{E} ( T_1\mathbbm{1}_{\{R_2=s\}}) 
	&=&	
			\xi_1 \biggl(	
		 	F_{S_{2,k}}(d_2)
		 	\overline{U}_{S_{1,k}}(s,d_1,Z(\beta_{2k}))	+ 
		 	\overline{U}_{S_{1,k}+S_{2,k} }(s,d_1,d_2,Z(\beta_{2k}))
		 	\biggr) \notag \\
		 	&& 
		 	+ \sum_{l=1}^{2k} \sum_{j_1,j_2,\ldots, j_l}  
			\xi_{j_1,j_2,\ldots,j_{l}} \prod_{m=1}^l \gamma_{j_m} 
			\biggl(
			F_{\tilde{S}_{2,k}}(d_2)
		 	\overline{U}_{\tilde{S}_{1,k}}(s,d_1,Z(\beta_{2k})) 
		 	\notag \\
		 	&& 
			+ \overline{U}_{\tilde{S}_{1,k} }(s,d_1,d_2,Z(\beta_{2k}))
			\biggr).
\EQN}
Therefore, refering to \eqref{eq:defTVaRAlloc1} (set $x_p:= VaR_p(R_2)$)
\gE{\BQNY
  TVaR_{p}(T_1,R_2)&=&
	\frac{1}{ 1-p} 
			\biggl[
			\xi_1 \biggl(
		 	F_{S_{2,k}}(d_2) \int_{x_p}^{\infty}	\overline{U}_{S_{1,k}}(s,d_1,Z(\beta_{2k})) ds +
		 	\int_{x_p}^{\infty} \overline{U}_{S_{1,k}}(s,d_1,d_2,Z(\beta_{2k})) ds
		 	\biggr) \notag\\ 
		 	&& 
		 	+ \sum_{l=1}^{2k} \sum_{j_1,j_2,\ldots, j_l}  
			\xi_{j_1,j_2,\ldots,j_{l}} \prod_{m=1}^l \gamma_{j_m} 
			\biggl(
		 	F_{\tilde{S}_{2,k}}(d_2)
		 	\int_{x_p}^{\infty} \overline{U}_{\tilde{S}_{1,k}}(s,d_1,Z(\beta_{2k})) ds
		 	\notag\\ 
		 	&& 
			+ \int_{x_p}^{\infty} \overline{U}_{\tilde{S}_{1,k} }(s,d_1,d_2,Z(\beta_{2k})) ds
			\biggr) 
			 \biggr].
\EQNY  }
Hence, the result follows easily. 
\QED\\
\proofprop{proposition:Value of the default option}
By definition
\BQNY 
	U(K)=\E{(R_2-K)_{+}} = \overline{F}_{R_2}(K) \E{R_2-K\vert R_2> K }.
\EQNY
Hence in view of Remark \ref{eq: Meanexcess_R2},  
\gE{\BQNY
	U(K)&=& 
			\xi_1 \biggl(
		 	F_{S_{1,k}}(d_1) \overline{U}_{S_{2,k}}(K,d_2,Z(\beta_{2k}))+	
		 	F_{S_{2,k}}(d_2)	\overline{U}_{S_{1,k}}(K,d_1,Z(\beta_{2k}))
		 	 +
		 	\overline{U}_{S_{1,k}+S_{2,k} }(K,d_1,d_2,Z(\beta_{2k}))
		 	\biggr) \notag\\ 
		 	&& 
		 	+ \sum_{l=1}^{2k} \sum_{j_1,j_2,\ldots, j_l}  
			\xi_{j_1,j_2,\ldots,j_{l}} \prod_{m=1}^l \gamma_{j_m} 
			\biggl(
			F_{\tilde{S}_{1,k}}(d_1)
			\overline{U}_{\tilde{S}_{2,k}}(K,d_2,Z(\beta_{2k}))
			\notag\\ 
		 	&& +
		 	F_{\tilde{S}_{2,k}}(d_2)
		 	\overline{U}_{\tilde{S}_{1,k}}(K,d_1,Z(\beta_{2k}))
			+ \overline{U}_{\tilde{S}_{1,k}+\tilde{S}_{2,k}}(K,d_1,d_2,Z(\beta_{2k}))
			\biggr) 
		 - K \overline{F}_{R_2}(K), 
\EQNY}
establishing the proof.
\QED\\
	\proofprop{proposition:unpaid excess losses}
The unpaid losses of the ceding insurer line of business is defined as follows 
	\BQNY 
		U(K_1,K)=\E{(T_1-K_1) \mathbbm{1}_{ \{R_2 > K \} }}=  \int_{K} ^ \infty \mathbb{E} ( T_1\mathbbm{1}_{\{R_2=s\}}) ds - K_1 \overline{F}_{R_2}(K) . 
	\EQNY
In light of \eqref{eq:CondExpT1TFinal}
	\gE{\BQNY
		U(K_1,K)&=&
   \xi_1 \biggl(
		 	F_{S_{2,k}}(d_2)	\overline{U}_{S_{1,k}}(K,d_1,Z(\beta_{2k}))+
		 	\overline{U}_{S_{1,k}}(K,d_1,d_2,Z(\beta_{2k}))
		 	\biggr) \notag\\ 
		 	&& 
		 	+ \sum_{l=1}^{2k} \sum_{j_1,j_2,\ldots, j_l}  
			\xi_{j_1,j_2,\ldots,j_{l}} \prod_{m=1}^l \gamma_{j_m} 
			\biggl(
		 	F_{\tilde{S}_{2,k}}(d_2)
		 	\overline{U}_{\tilde{S}_{1,k}}(K,d_1,Z(\beta_{2k}))
		 	\notag\\ 
		 	&& 
			+ \overline{U}_{\tilde{S}_{1,k}}(K,d_1,d_2,Z(\beta_{2k}))
			\biggr) -
			K_1 \overline{F}_{R_2}(K).
	\EQNY }
Hence the proof is complete. 
\QED
\appendixtitleon
\appendixtitletocon
\begin{appendices}
			     	\section{Properties of mixed Erlang distribution}
\begin{lem}\label{lem:pdfStop_Loss}
For a deductible $d >0$, if $X \sim ME(\beta, \utilde{Q})$  then the df of  $Y:=(X-d)_+$ is given by 
\BQN \label{eq:pdfStop_Loss}
F_Y(y) = 
\left\{
			 	\begin{array}{lcl}
         			F_X(d)   & \mbox{for} & y=0, \\
         			F_X(y+d)   &\mbox{for} & y>0,
              	\end{array}
        	\right.
\EQN
where 
$$F_X(y+d)
	= 
	\sum_{k=0}^\infty 
	\Delta_{k}(d,\beta,\utilde{Q} )
	W_{k+1}(y,\beta),$$
with
\BQN \label{eq:ceoffGamma}
\Delta_{k} (d,\beta,\utilde{Q}) = \frac{1}{\beta} \sum_{j=0}^\infty 
	q_{j+k+1}w_{j+1}(d,\beta).
\EQN 
\end{lem}
\COM{
\BRM \label{rem:momentStopLoss}
It follows that the $h$-th moment of the stop loss risk $Y$ can be expressed as
\BQNY \label{eq:momentStop_Loss}
	\mu^h(d,\utilde{Q}) =\E{(X-d)^h _+ } 
	= \sum_{k=0}^\infty 
	\Delta_k(d,\beta) 
	\int_{d}^\infty y^h w_{k+1}(y,\beta) dy
	=\beta^{-h} \sum_{k=0}^\infty 
	\Delta_k(d,\beta,\utilde{Q}) \frac{\Gamma(h+k +1)}{\Gamma(k + 1 )},
\EQNY

with $\Gamma(.) $ the Gamma function.
\ERM}
\begin{lem}\label{lem:Aggreg_Stop_Loss_Indep}
	Let $X_1$  and $X_2$ be two independent risks such that  $X_i\sim ME(\beta,\utilde{Q}_i), i=1,2$. If $d_i, i=1,2$ are positive then  $R_2= Y_1+ Y_2 $, with $Y_i=(X_i-d_i)_+ , i=1,2$, has df 
	\BQN \label{eq:pdfStop_Loss}
F_{R_2}(s) = 
\left\{
			 	\begin{array}{lcl}
         			F_{X_1}(d_1) F_{X_2}(d_2)   & \mbox{for} & s=0 ,\\
         			F_{X_1}(d_1) F_{X_2}(s+d_2)
         			+ F_{X_2}(d_2) F_{X_1}(s+d_1)
         			+ F_{X1+X_2}(d_1,d_2,s)				  &\mbox{for} & s>0,
              	\end{array}
        	\right.
\EQN
	where
	\BQNY
		F_{X1+X_2}(d_1,d_2,s)= \sum_ {k=0}^{\infty}   \sum_ {j=0}^{\infty}	
					\Delta_k(d_1,\beta,\utilde{Q}_1) \Delta_j(d_2,\beta,\utilde{Q}_2)   W_{k+j+2}(s,\beta).
	\EQNY
\BRM
Given the tractable 
 of the df of $R_2$, the VaR of $R_2$ at a confidence level of $p \in (0,1)$ is the solution of 
$$
         			F_{X_1}(d_1) F_{X_2}(d_2)+ F_{X_1}(d_1) F_{X_2}(VaR_{R_2} (p) +d_2)
         			+ F_{X_2}(d_2) F_{X_1}(VaR_{R_2} (p)+d_1)
         			+ F_{T_2}(VaR_{R_2} ( p))	 = p ,$$
 which can be solved numerically.
 In addition,  the TVaR of $R_2$ at a confidence level $p \in (0, 1)$ is given by (set $x_p:=VaR_{R_2} (p)$)
 \BQNY
 TVaR_{R_2}(p) &=&
 \frac{\beta^{-1}}{1-p}
	\biggl(
  F_{X_1}(d_1) 
  \overline{U}_{X_1}(x_p,d_1,\beta)
	+ F_{X_2}(d_2)  
	\overline{U} _{X_2}(x_p,d_2,\beta)
	+
	\overline{U}_{X_1+X_2}(x_p,d_1,d_2,\beta) \biggr),
\EQNY
where
\BQNY
&&\overline{U} _{X_i}(x_p,d,\beta)
	= 
	 \sum_ {k=0}^{\infty} 
			 (k+1)   \Delta_k(d_i,\beta,\utilde{Q}_i)
			 \overline{W}_{k+2}(x_p,\beta),\\
	&&\overline{U}_{X_1+X_2}(x_p,d_1,d_2,\beta)= 
		\sum_ {k=0}^{\infty}   \sum_ {j=0}^{\infty} (k+j+2)	
		\Delta_k(d_1,\beta,\utilde{Q}_1) \Delta_j(d_2,\beta,\utilde{Q}_2)   W_{k+j+3}(x_p,\beta).
		\EQNY 
\COM{ Furthermore,  the $h$-th moment of the aggregated stop loss risk  has a closed form as follows
\BQNY \label{eq:momentStop_Loss_Aggreg}
	\E{R_2 ^h  } 
	= 	F_{X_1}(d_1) \mu^h (d_2, \utilde{Q}_2)  + F_{X_2}(d_2)  \mu^h (d_1, \utilde{Q}_1)
	 + 
	 \beta^{-h} \sum_ {k=0}^{\infty}
	 \sum_ {j=0}^{\infty}	
	\Delta_k(d_1,\beta,\utilde{Q}_1)
	\Delta_j(d_2,\beta,\utilde{Q}_2)
	\frac{\Gamma(h+k + j +2)}{\Gamma(k + j + 2 )}.
\EQNY}
\ERM
\end{lem}	
\textbf{Proof.}		
Since $Y_1$ and  $Y_2$ are independent risks which have mixed distribution, the df of $R_2$  can also be expressed as a df  of a mixed distribution which depends on the value of  $s$ as follows:
\BIT
 \item the discrete part of $F_{R_2}$ is obtained for $s=0$, specifically we have 
 \BQN \label{eq: dis_part_FR2}
 		F_{R_2}(0) = \pk{Y_1 + Y_2 \leqslant 0} = \pk{Y_1 + Y_2 = 0} = \pk{Y_1=0, Y_2 = 0} = F_{X_1}(d_1) F_{X_2}(d_2),
 \EQN
 \item for $s>0$ the continious part of $F_{R_2}$ is given by 
 \BQN \label{eq:Cont_Part_FR_2}
 		F_{R_2}(s) &=& \pk{Y_1 + Y_2 \leqslant s } \notag \\
 				   &=& 
 				   		\pk{ Y_1 + Y_2 \leqslant s, Y_1 = 0, 0< Y_2  \leqslant s} 
 				   		+ \pk{ Y_1 + Y_2 \leqslant s,  0< Y_1  \leqslant s, Y_2 = 0} \notag \\
 				   &&		
 				   		+ \pk{ Y_1 + Y_2 \leqslant s , 0< Y_1  \leqslant s, 0< Y_2  \leqslant s} \notag \\
 				   &=&
 				   	\pk{  Y_1 = 0,  0 <Y_2  \leqslant s} 
 				   		+ \pk{ 0 < Y_1  \leqslant s, Y_2 = 0} 
 				   		+ \pk{ Y_1 + Y_2 \leqslant s , 0< Y_1  \leqslant s, 0< Y_2  \leqslant s} \notag \\
 				   	&=&
 				   	F_{X_1}(d_1) F_{X_2}(s+d_2)
         			+ F_{X_2}(d_2) F_{X_1}(s+d_1)
         			+ \int_0^s F_{X_1}( s- u + d_1) f_{X_2}(u+ d_2) du. 	
 \EQN
 Let  $F_{T_2}(s) : = \int_0^s F_{X_1}( s- u + d_1) f_{X_2}(u+ d_2) du $, by Lemma \ref{lem:pdfStop_Loss}
this can be written as  
  \BQN \label{eq:convY1Y2}
  		F_{T_2}(s)
  					= \sum_ {k=0}^{\infty}
	 \sum_ {j=0}^{\infty}	
	\Delta_k(d_1,\beta_1)
	\Delta_j(d_2,\beta_2)
	\int_0^s 
	W_{k+1}(s- u,\beta) 
	w_{j+1}(u,\beta) du.
  \EQN
\EIT
It can be seen that
$\int_0^s 
	W_{k+1}(s- u,\beta) 
	w_{j+1}(u,\beta) du$
is a convolution of two independent Erlang risks with a common scale parameter $\beta$, which is again an Erlang risk with shape parameter $k+j+2$ and scale parameter   $\beta$. Thus combining \eqref{eq: dis_part_FR2}, \eqref{eq:Cont_Part_FR_2} and \eqref{eq:convY1Y2} the claim follows easily. 
\begin{lem}\label{lem:tranformation ME}
 Let $X \sim ME(\beta, \utilde{Q})$ with pdf $f(x,\beta,\utilde{Q}) $, if $g$ is some \aH{positive} function such that $\E{g(X)}<\infty$,  then $c(x,\beta,\utilde{Q})=\frac{g(x)f(x,\beta,\utilde{Q})}{\E{g(X)}}$ is again a pdf of mixed Erlang distribution with scale parameter $Z(\beta)$ and mixing weights $\utilde{\Theta}(\utilde{Q})=(\theta_1,\theta_2,\ldots)$, with 
$$c(x,\beta,\utilde{Q})=\sum_{k=1}^\infty \theta_k w_k(x,Z(\beta)),$$
where
\BIT
	\item  $Z(\beta)=2 \beta$ and
	$ \theta_k = \frac{1}{2^{k-1}} 
	\sum_{j=1}^k 
	\begin{pmatrix}
 				k-1\\
				j-1
	\end{pmatrix}
	q_j \sum_{l=k-j+1}^\infty q_l ,$
	for $g(x)=2\overline{F}(x)$,
	\item  $Z(\beta)= \beta$ and
	\BQNY
\theta_k=
   \left\{
			 	\begin{array}{lcl}
         			0   & \mbox{for} & k \leqslant t , \notag \\
         			\frac{q_{k-t}\frac{\Gamma(k)}{\Gamma(k-t)}}{\sum_{j=1}^\infty q_j\frac{\Gamma(j+t)}{\Gamma(j)} }   &\mbox{for} & k > t,
              	\end{array}
        	\right.
\EQNY
	 for $g(x)=x^t$ with $t \in \R$,
	 
	\item  $Z(\beta)= \beta + t $ and
	$ \theta_k = \frac{q_k \overline{\beta}^k}{\sum_{j=1}^\infty q_j\overline{\beta}^j } 
	 $ with $\overline{\beta} =\frac{\beta}{\beta + t} ,$
	 for $g(x)=e^{-tx}$ with $t \in \N$.
		
\EIT
\end{lem}	
\textbf{Proof.} 
We have 
\BQN \label{eq:def1}
 c(x,\beta,\utilde{Q})=\frac{g(x)f(x,\beta,\utilde{Q})}{\E{g(X)}} =  
 \frac{1}{\E{g(X)}}
 \sum_{k=1}^{\infty} q_k  \frac{\beta^k}{(k-1)!} g(x) x^{k-1} e^{-\beta x}.
\EQN
For $g(x)= x^t$ one can write \eqref{eq:def1} as follows 
\BQNY
 c(x,\beta,\utilde{Q}) &= & 
 \frac{1}{\E{X^t}}
 \sum_{k=1}^{\infty} q_k  \frac{\beta^k}{(k-1)!}  x^{t+k-1} e^{-\beta x} \\
 &=&\sum_{k=1}^{\infty} \biggl(\frac{q_{k}\frac{\Gamma(k+t)}{\Gamma(k)}}{\sum_{j=1}^\infty q_j\frac{\Gamma(j+t)}{\Gamma(j)} }\biggl) w_{k+t}(x,\beta)\\
 &= & 
 \sum_{s=t+1}^{\infty} \biggl(\frac{q_{s-t}\frac{\Gamma(s)}{\Gamma(s-t)}}{\sum_{j=1}^\infty q_j\frac{\Gamma(j+t)}{\Gamma(j)} }\biggl) w_{s}(x,\beta)
 \\
 &= & 
 \sum_{s=1}^{\infty} \theta_s w_{s}(x,\beta),
\EQNY
with 
\BQNY
		\theta_s
		&=&
	       \left\{
			 	\begin{array}{lcl}
         			0   & \mbox{for} & s \leqslant t, \notag \\
         			\frac{q_{s-t}\frac{\Gamma(s)}{\Gamma(s-t)}}{\sum_{j=1}^\infty q_j\frac{\Gamma(j+t)}{\Gamma(j)} }  &\mbox{for} & s < t.
              	\end{array}
        	\right.  
        	\\ 
\EQNY
For $g(x)= e^{-tx}$, \eqref{eq:def1} can be expressed as follows (set $\overline{\beta} := \frac{\beta}{\beta+t}$)
\BQNY
 c(x,\beta,\utilde{Q}) &= & 
 \frac{1}{\E{e^{-tX}}} 
 \sum_{k=1}^{\infty} q_k
 \frac{\beta^k}{(k-1)!}  x^{k-1} e^{-(\beta + t) x} \\
  &= &  
  \sum_{k=1}^{\infty} 
  \Biggl(
  \frac{q_k \overline{\beta}^k}{\sum_{j=1}^\infty q_j\overline{\beta}^j }
  \Biggr)
  w_k(x,\beta +t) \\
  &= & 
  \sum_{k=1}^{\infty} 
  \theta_k w_k(x,\beta +t). 
\EQNY
	For $g(x)=2\overline{F}(x)$, see \cite{Cossette_al13} for the proof.

The results presented in the next two lemmas can be found in Section 2.2  of \cite{Willmot_Woo97} and Section 7.2 of \cite{Lee_Lin10}, respectively.
\begin{lem}\label{lem:pdfMErlang}
If $X \sim ME(\beta_1,\utilde{Q})$, then for any positive constant $\beta_2 \ge \beta_1$ we have
\BQNY
		X \sim ME(\beta_2, \   \utilde{\Psi}(\utilde{Q})), \quad \utilde{\Psi}(\utilde{Q})=(\psi_1, \psi_2, \ldots),
\EQNY
where 
		\BQNY
		\psi_k=\sum_ {i=1}^{k} q_i 		
		 	\begin{pmatrix}
 				k-1\\
				i-1
			\end{pmatrix}		
 		\left(\frac{\beta_1}{\beta_2}\right)^i \left(1-\frac{\beta_1}{\beta_2}
 		\right)^{k-i}, \quad k\ge 1.
	\EQNY	
\end{lem}
%
%
\begin{lem}\label{lem:Aggregation}
	Let $X_1,X_2$ be two independent random variables such that $X_i\sim ME(\beta, \utilde{Q}_i)  ,i=1,2$, then $S_2=X_1+X_2\sim ME(\beta, \utilde{\Pi}\{\utilde{Q}_1, \utilde{Q}_2\})$ with 
	\BQNY
		\pi_l\{\utilde{Q}_1, \utilde{Q}_2\}=
			\left\{
			 	\begin{array}{rcl}
         			0  & \mbox{for} & l=1, \\
         			\sum_ {j=1}^{l-1} q_{1,j} \ q_{2,l-j}& 		    \mbox{for} & l>1.
              	\end{array}
        	\right.
	\EQNY		
\BRM \label{rem:AggregationN}
According to Cossette et al. (2012) (Remark 2.1), the results in Lemma \ref{lem:Aggregation} can be extended to $S_n= \sum_{i=1}^n X_i$, provided that $X_1,\ldots,X_n$ are independent, $X_i\sim ME(\beta, \utilde{Q}_i) $  for $i=1,\ldots,n$. Specifically, $S_n \sim ME(\beta, \utilde{\Pi}\{\utilde{Q}_1,\ldots, \utilde{Q}_n\})$ where the individual mixing probabilities can be evaluated iteratively as follows
\BQNY
		\pi_l \{\utilde{Q}_1,\ldots, \utilde{Q}_{n+1}\}=
			\left\{
			 	\begin{array}{rcl}
         			0  & \mbox{for} & l=1,\ldots,n, \\
         			\sum_ {j=n}^{l-1} \pi_j\{\utilde{Q}_1,\ldots, \utilde{Q}_{n}\} \ q_{n+1,l-j}& 		    \mbox{for} & l=n+1,n+2,\ldots.
              	\end{array}
        	\right.
	\EQNY
\ERM
\end{lem}	
	\section{ Joint density of sums of Sarmanov random vectors}
One of the main features of the Sarmanov distribution is that its pdf can be used to derive some results in analytical way. For instance Vernic \cite{Vernic_2014}  have derived general formula for the  density of the sum of several rv joined by the Sarmanov distribution.
Below  we derive the joint density of $n$ random vectors where each vector consists of $k$ elements and we denote the sum of elements within each random vector  as 
$
 S_i:=\sum_{j=(i-1)k+1}^{ik}  X_j, i=1,\ldots, n.
$
 Furthermore, we assume that the joint distribution of the overall random vectors $(X_1, \ldots,X_{nk})$ has the Sarmanov distribution \gE{with any kernel function satisfying \eqref{eq:Condition_pdfCopulaN}.}
  \begin{theo} \label{theorem:JointVectors}
 The joint density  of $(S_1,\ldots, S_n)$ is given by
\BQNY
 \gE{\zeta}(u_1,\ldots,u_n)   &=&	
			\prod_{i=1}^{n} f_{S_i}(u_i) + \sum_{h=2}^{n} \sum_{1 \leqslant j_1 < j_2< \ldots < j_h \leqslant n } 
				\alpha_{ j_1, \ldots ,j_h} 
				\prod_{i=1}^{n} \gE{f^{(*)}_{S_i}(u_i)},
\EQNY
where 
$$f_{S_i}(u_i) =( f_{{(i-1)k+1}} * \ldots * f_{ik}) (u_i), $$
 \gE{$$f^{(*)}_{S_i}(u_i) = ( f_{{(i-1)k+1}} ^{(*)} * \ldots * f_{ik}^{(*)}) (u_i) ,$$}
 with
$$\gE{f_m^{(*)}(x_m)} =  \left\{
			 	   \begin{array}{lcl}
         			 f_m(x_m) & \mbox{if} & m \notin \{ j_1,j_2,\ldots, j_h \}, \\
         		    \phi_{x_m}(s) f_m(x_m) ds & \mbox{if} & m \in \{j_1,j_2,\ldots, j_h\},m=1,\ldots,{nk}.
              	\end{array}
      \right.$$
\end{theo} 	
\textbf{Proof.}
\gE{
The joint density of  $(S_1,\ldots, S_n)$ is determined in term of the joint density  of  $(X_1,\ldots,X_{nk})$ as follows}
\BQN \label{eq:defN}
			\gE{\zeta}(u_1,\ldots,u_n)=
	 		\int \ldots \int_{s_1 = u_1,s_2   = u_2,\ldots, s_n   =   u_n}  h(\x)dx_1\ldots dx_{nk-1},
\EQN
with $\x=(x_1,\ldots,x_{nk}), 
 s_1= x_1+\ldots + x_k,
 s_2 =  x_{k+1} +\ldots+X_{2k},
 s_n = x_{nk-k+1} + \ldots + x_{nk}.
$\\
Refering to \eqref{eq:pdfCopulaN},  
\BQNY
	h(\x)&=& 
			\prod_{i=1}^{nk}  f_i(x_i)  \biggl(
				1+ \sum_{h=2}^{n} \sum_{1 \leqslant j_1 < j_2< \ldots < j_h \leqslant n } 
				\alpha_{ j_1, \ldots ,j_h}
				\prod_{k=1}^h \phi_{j_k}(x_{j_k})
			\biggr)\\
		 &=& 
			\prod_{i=1}^{nk}  f_i(x_i) 
				+ \sum_{h=2}^{n} \sum_{1 \leqslant j_1 < j_2< \ldots < j_h \leqslant n } 
				\alpha_{ j_1, \ldots ,j_h}
				\prod_{k=1}^h \phi_{j_k}(x_{j_k}) f_{j_k}(x_{j_k})
				\prod_{m \notin \{j_1,j_2,\ldots, j_h\} }f_m(x_m)	\\
			 &=& 
			\prod_{i=1}^{nk}  f_m(x_m) 
				+ \sum_{h=2}^{n} \sum_{1 \leqslant j_1 < j_2< \ldots < j_h \leqslant n } 
				\alpha_{ j_1, \ldots ,j_h} 
				\prod_{m=1}^{nk} \gE{f_m^{(*)}(x_m)}	,		
\EQNY
where
for $m=1,\ldots,{nk}$
$$\gE{f_m^{(*)}(x_m)} =  \left\{
			 	   \begin{array}{lcl}
         			 f_m(x_m) & \mbox{if} & m \notin \{ j_1,j_2,\ldots, j_h \}, \\
         			  \phi_{m}(x_{m}) f_m(x_m) & \mbox{if} & m \in \{j_1,j_2,\ldots, j_h\}.
              	\end{array}
      \right.$$
Therefore, one can express \eqref{eq:defN} as a sum of convolutions 	as follows (set $\x _{ik}:= (x_{(i-1)k+1}, \ldots, x_{ik-1}), i=1,\ldots, n$)

\BQNY
			\gE{\zeta}(u_1,\ldots,u_n)
			&=&	
			\prod_{i=1}^{n} \int_{\R^{k-1}} 
			\int  \prod_{m=(i-1)k+1}^{ik-1}  f_j(x_j)  f_{ik}(u_i - \sum_{m=(i-1)k+1}^{ik-1}  x_j)
			d\x _{ik} 
			\\
			&+&
			 \sum_{h=2}^{n} \sum_{1 \leqslant j_1 < j_2< \ldots < j_h \leqslant n } 
				\alpha_{ j_1, \ldots ,j_h} \\
			&&
				\prod_{i=1}^{n}  \int_{\R^{k-1}} \int
				 \prod_{m=(i-1)k+1}^{ik-1}\gE{f_m^{(*)}(x_m)}
				\gE{f_{ik}^{(*)}(u_j - \sum_{m=(i-1)k+1}^{ik-1}  x_m)d\x _{ik}}\\
			&=&	
			\prod_{i=1}^{n} f_{S_i}(u_i) + \sum_{h=2}^{n} \sum_{1 \leqslant j_1 < j_2< \ldots < j_h \leqslant n } 
				\alpha_{ j_1, \ldots ,j_h} 
				\prod_{i=1}^{n} \gE{f^{(*)}_{S_i}(u_i).}	
\EQNY
\end {appendices}
\textbf{Acknowledgments}. The author acknowledges partial financial support received from the project RARE -318984
 (an FP7  Marie Curie IRSES Fellowship) and Vaudoise Assurances. 	
 \bibliographystyle{plain}
\bibliography{aggSarmanovD}
\end{document}